\documentclass[letterpaper,11pt]{article}
\usepackage{jheppub}

\usepackage[utf8]{inputenc}

\usepackage{braket}
\usepackage{tikz}
\usepackage{amssymb}
\usepackage{amsmath}
\usepackage{color}
\usepackage{mathrsfs}
\usepackage{graphicx}
\usepackage{tensor}
\usepackage{subcaption}
\usepackage[labelfont={small, bf, sf}, textfont={small,sf}]{caption}
\usepackage{ragged2e} % needed for JHEPthesis bibtex style file
\usepackage[colorlinks=true]{hyperref}

\newsavebox{\largestimage}

\newcommand{\scri}{\mathscr{I}}

\newcommand{\lie}{\pounds}
\newcommand{\cu}{\mathcal{U}}

\DeclareMathOperator{\diag}{diag}
\DeclareMathOperator{\sgn}{sgn}

\newcommand{\beq}{\begin{equation}}
\newcommand{\eeq}{\end{equation}}

\begin{document}

\title{Cosmic footballs from superrotations}

\author[a,b]{Eugene Adjei,}
\author[a]{William Donnelly,}
\author[a,c]{Victor Py,}
\author[a]{Antony J. Speranza}

\affiliation[a]{Perimeter Institute for Theoretical Physics, 31 Caroline St. N, N2L 2Y5, Waterloo ON, Canada}
\affiliation[b]{Department of Physics and Astronomy, 
University of Waterloo, 200 University Ave W, Waterloo ON,
Canada}
\affiliation[c]{Department of Mathematics, University of California, Davis, CA 95616, USA} 

\emailAdd{eadjei@perimeterinstitute.ca}
\emailAdd{wdonnelly@perimeterinstitute.ca}
\emailAdd{vpy@ucdavis.edu}
\emailAdd{asperanz@gmail.com}

\abstract{
Superrotations arise from singular vector fields on the celestial sphere in 
asymptotically flat space, and their finite integrated versions have been 
argued by Strominger and Zhiboedov to insert cosmic strings into the spacetime.
In this work, we argue for an alternative definition of the action of superrotations on 
Minkowski space that avoids introducing any defects.  This involves realizing the 
finite superrotation not as a diffeomorphism between spaces, but as a mapping of
Minkowski space to itself that may be multivalued or non-surjective.  This eliminates 
any defects in the bulk spacetime at the expense of allowing for defects in the boundary
celestial sphere metric.  We further explore the geometry of the spatial surfaces 
in the superrotated spaces, and note that they intersect null infinity at the 
singularity of the superrotation, causing a breakdown in the large $r$ asymptotic
expansion there.  To determine how these surfaces embed into Minkowski space, 
a derivation of the finite superrotation transformation is presented
in both Bondi and Newman-Unti gauges.  The latter is particularly interesting, since the superrotations
are shown to preserve the hyperbolic slicing of Minkowski space in Newman-Unti gauge,
and this gauge 
also provides a means for extending the geometry beyond the Bondi coordinate patch.
We argue that the new interpretation for the action of superrotations on spacetime
motivates consideration of a wider class of celestial sphere metrics and asymptotic 
symmetry groups.
}
\maketitle

\section{Introduction}

The symmetries of general relativity in asymptotically flat space at null infinity
comprise an infinite-dim\-en\-sion\-al enlargement of the Poincar\'e group in four dimensions, 
known as the BMS group \cite{Bondi1962, Sachs1962b, Sachs1962a}.  
In addition to the expected Lorentz 
transformations, the BMS group also contains a subgroup of  supertranslations, which act as
angle-dependent translations along null infinity. The supertranslations are  parameterized
by functions on the celestial 2-sphere, and the Lorentz group acts on them via its natural
action as the conformal isometry group of $S^2$.  
Infinitesimally, the 2-sphere possesses a much larger conformal isometry algebra that includes
generators with pole singularities, and it has been suggested that this extended
algebra, together with the supertranslations, should constitute the true symmetry
algebra for asymptotically flat general relativity at null infinity
\cite{Barnich2010c, Barnich2011b, Barnich2011a, Banks:2003vp}.  
The additional generators in the extended BMS group have 
been dubbed ``superrotations,'' although since they can be thought of as 
local generalizations of the boosts of Minkowski space, they might more appropriately be 
called ``superboosts'' \cite{Compere2018}.

The singular nature of  superrotation generators raises questions about their status as 
true symmetries of general relativity.  On the one hand, postulating superrotation symmetry of the 
gravitational S-matrix leads to a derivation of the subleading soft graviton theorem as a 
Ward
identity \cite{Cachazo2014, Kapec2014},
similar to how the leading soft graviton theorem follows from invariance under 
supertranslations \cite{Strominger2013, He2014a}.  
Since the subleading soft theorem can be derived using independent arguments
\cite{Broedel2014, Bern2014}, this 
lends support to superrotations being symmetries of the theory.  Another argument
in favor of physicality of superrotations is the close relation between this symmetry and 
novel types of gravitational memory \cite{Pasterski2016, Flanagan:2015pxa}. 
On the other hand, the singular nature of 
the superrotation transformations makes it unclear how they act on spacetime,
and calls into question whether they can be integrated to finite, smooth diffeomorphisms.  
Investigations into the finite superrotation transformations have suggested that 
their action on spacetime is not smooth, but rather inserts defects such as cosmic 
strings
\cite{Strominger:2016wns, Compere:2016jwb, Capone:2019aiy}.  Although this action of 
the finite superrotations is well-defined, the fact that it injects matter in the form of a 
cosmic string 
raises doubts about its interpretation as a symmetry transformation for the gravitational theory.  

The purpose of the present work is to argue for a different perspective on how superrotations
act on spacetime.  In \cite{Strominger:2016wns, Compere:2016jwb}, 
the finite superrotations act in such a 
way that the surfaces at constant radius $r$ and retarded time $u$ are regular in the limit
$r\rightarrow \infty$.  
Although these surfaces have punctures located at the poles of the superrotation
transformation, the punctures are removable at infinite $r$ in the sense that the 
instrinsic metric on
the surfaces is precisely equal the the round sphere metric almost everywhere, except at the 
measure zero set of puncture points.  We argue in section \ref{sec:defect} that the
regularity requirement of these celestial 2-spheres is responsible for the presence of 
defects such as cosmic strings in the superrotated bulk.  
By instead imposing that the celestial sphere develop a conical defect 
due to the superrotation, one can eliminate the bulk cosmic string defect.  
We conclude that superrotations should be defined to maintain regularity
in the bulk, at the expense of changing the metric on the celestial sphere by allowing 
for non-removable defects.

The cosmic string spacetimes arise from   
superrotations that act discontinuously on the celestial $2$-sphere.  
They are produced by the transformation $z \to G(z) = z^\alpha$, 
which has a branch discontinuity along the negative real axis.  
This discontinuity extends to the bulk transformation 
of Minkowski space, and leads to a map that either overwraps the space as a multivalued 
transformation, or is not surjective, mapping Minkowski space to a proper subset.  
If one erroneously identifies two points that originated on either side of the branch after applying the diffeomorphism, one would conclude that the bulk spacetime must have developed a cosmic string defect due to the angular excess or deficit.  
However, by making the correct identifications of coincident points following the singular diffeomorphism, 
one finds a flat metric devoid of any cosmic strings.  

Besides the subtleties associated with identifying the proper coordinate ranges, the resulting
metric has a number of unintuitive features that arise due to the singular nature of the branch
point.  
After applying the $z^\alpha$ superrotation to Minkowski space, the metric is given by
equation
(\ref{eqn:zalphamet}).  
This metric is locally flat and formally can be written as an asymptotic expansion 
in powers of $\frac{1}{r}$.  However, the singularity in the superrotation means that this 
asymptotic expansion breaks down near the poles of the transformation, with subleading terms in 
$\frac1r$ dominating over the leading terms.  In fact, the surfaces at constant $(u,r)$ behave quite 
differently from spheres.  In certain regions, there is an infinite distance on these surfaces
between an arbitrary point and the pole at $z=0$, suggesting a shape more like an (American) football,
becoming long and narrow near the poles
(see figure \ref{fig::embedR3pos}) .  Other regions are even 
more bizarre, with the length to the pole being finite,
while the circumference of a circle surrounding pole diverging as $z\rightarrow 0$.  This implies a
highly negatively curved space near the pole, giving the $(u,r)$ surface a horn-like shape
(figure \ref{fig::embedR3neg}), and 
also implies the existence of a curvature singularity at finite distance.

To better understand the regular action of superrotations on Minkowski space, in section
\ref{section:superrotation} we explicitly construct the transformation 
between Minkowski spacetime in standard retarded coordinates 
and the superrotated spacetime of Comp\`ere and Long \cite{Compere:2016jwb}.
This is done by solving for the transformation perturbatively near $\scri^+$, 
and then using the fact that lines of constant $u, z, \bar z$ in Bondi coordinates are 
null geodesics to solve for the flow  everywhere in the bulk.  The details of this derivation
are presented in appendix \ref{sec::diffeo}, which gives an alternative 
to the original derivation of the finite 
superrotation given in \cite{Compere2016b}, and the transformation
is presented in equations (\ref{eqn:nullgeo}) and (\ref{eqn:finitesuperrot}). 
We also derive the transformation in Newman-Unti gauge, which differs 
from Bondi gauge in the choice of radial coordinate \cite{Newman:1962cia}.  
The Newman-Unti superrotated metric (\ref{eqn:gNUcomp}) 
takes a much simpler form than the Bondi gauge metric, and the superrotation in Newman-Unti
gauge has the interesting feature of preserving hyperboloids at fixed 
Lorentzian distance from the origin. Using these 
formulas, we can show how the cosmic footballs and cosmic horns embed into Minkowski
space.  As argued from analysis of the intrinsic geometry, the surfaces intersect asymptotic
infinity at the poles of the transformation, and, depending on the values of $u$ and $\alpha$, 
can approach the celestial sphere at $u=0$, timelike infinity, or spacelike infinity.  
We also plot the $r=0$ hypersurface in 
figure \ref{fig:r=0surf}, which clearly indicates the existence of a nonempty interior, and 
discuss how to extend the metric into this region using Newman-Unti coordinates.

This work concludes with section \ref{sec:discussion} 
with a discussion of our findings and their implications for 
the degrees of freedom in asymptotically flat general relativity.

\section{Conical defect in the finite superrotation}
\label{sec:defect}

Superrotations were proposed in 
\cite{Barnich2010c, Banks:2003vp}  as an infinite-dimensional extension of 
the asymptotic symmetry group of asymptotically flat general relativity.  They arise 
as generalizations of Lorentz transformations, and are characterized by their action
on the celestial 2-sphere.  The Lorentz group acts as the global conformal
isometry group of the celestial sphere, while the superrotations act as local conformal
isometries, which may not be globally defined due to singular points.  

This work deals with the action of superrotations on Minkowski space, the vacuum for 
asymptotically flat general relativity.  Their action on all of Minkowski space can be derived 
from their action on the celestial sphere by demanding that the resulting metric obey the 
Bondi gauge conditions \cite{Bondi1962, Madler2016}.
The Bondi coordinate system involves a retarded time variable $u$, a radial coordinate $r$,
and spatial coordinates $z^A = (z,\bar{z})$.  The radial generator $\partial_r^a$ is required 
to be tangent to null geodesics, and the coordinate $r$ is adjusted so that the metric determinant 
on surfaces of constant $(u,r)$ is equal to $-r^4 \gamma(z,\bar z)^2$, where $-\gamma(z,\bar z)^2$ 
is the unit 
sphere metric determinant in the $(z,\bar z)$ coordinates.  
Any asymptotically flat metric 
in this coordinate system can be expressed as \cite{Flanagan:2015pxa} 
\beq \label{eqn:ds2bondi}
ds^2 = -U e^{2\beta} du^2 -2 e^{2\beta} du\,dr + q_{AB}(dz^A - \cu^A du)(dz^B-\cu^B du),
\eeq
with the following falloff conditions at large $r$,
\begin{subequations}
\begin{align}
U &= 1 - \frac{2m}{r} +\mathcal{O}(r^{-2}) \\
\beta &= \mathcal{O}(r^{-1}) \\
q_{AB} &= r^2 \gamma_{AB} + \mathcal{O}(r) \\
\cu^A &= \mathcal{O}(r^{-2}) .
\end{align}
\end{subequations}
Here, $\gamma_{AB}$ defines the metric on the unit sphere, which in the $(z,\bar z)$ coordinate
system is given by
\beq \label{eqn:gzbarz}
\gamma_{AB}dz^A dz^B = 2 \gamma(z,\bar z)dz d\bar z, 
\qquad \gamma(z,\bar z) = \frac{2}{(1+ z\bar z)^2}.
\eeq

Minkowski space in standard retarded coordinates is of this form, with $U = 1$, $\beta = 0$, $\cu^A = 0$,
and $q_{AB} = r^2\gamma_{AB}$.  Superrotations define a class of transformations that preserve
the form of the metric (\ref{eqn:ds2bondi}), but yield different values for $U$, $\beta$, $\cu^A$, and 
$q_{AB}$.  Each finite superrotation is defined by a locally holomorphic function $G(z)$ on the 
2-sphere, and this produces a mapping of Minkowski space to itself.  This mapping was first
derived in \cite{Compere2016b}, and an alternate
derivation is discussed in section \ref{section:superrotation} and 
appendix \ref{sec::diffeo} of the present paper.  
Comp\`ere and Long derived the result of pulling back the 
standard Minkowski metric by this transformation, which is also a metric of the form
(\ref{eqn:ds2bondi}) with \cite{Compere:2016jwb}
\begin{subequations}
\label{eqn:gCL}
\begin{align}
g_{uu} &=-U e^{2\beta}+ q_{AB}\cu^A\cu^B  = -1 - \frac{2 u V}{\sqrt{r^2+u^2 V}}\label{eqn:Ue2b} \\
g_{ur} &= -e^{2\beta} = \frac{-r}{\sqrt{r^2+u^2V}} \\
g_{uA} & = -q_{AB} \cu^B = -D_A\sqrt{r^2 + u^2 V} \\ 
g_{AB} &= q_{AB} = (r^2 + 2 u^2 V) \gamma_{AB} + 2 u\sqrt{r^2 + u^2 V} \,T_{AB},
\label{eqn:qAB}
\end{align}
\end{subequations}
where the components of $T_{AB}$ are given by 
\beq\label{eqn:Tzz}
T_{zz} = -\frac12 \{ G(z),z\}, \quad T_{\bar z\bar z} = -\frac12\{ \bar G(\bar z) ,\bar z\}, \quad
T_{z\bar z} = 0,
\eeq
$\{G(z),z\} 
= \partial_z \left(\frac{\partial^2_z G}{\partial_z G}\right) - 
\frac12\left(\frac{\partial_z^2 G}{\partial_z G}\right)^2$ is the Schwarzian derivative, and 
\beq
\label{eqn:V}
V = \frac12 \gamma^{AB} \gamma^{CD}T_{AC} T_{BD} 
= \frac{\{G,z\} \{\bar G,\bar z\}}{4\gamma(z,\bar z)^2}.
\eeq

Except for 
global conformal transformations of the form $G(z) = \frac{a z + b}{c z + d}$,
a generic holomorphic map $G(z)$ 
contains points where the Schwarzian diverges, and may also have branch 
cuts across which $G(z)$ is discontinuous.  A specific class of maps of this type were considered by Strominger
 and Zhiboedov \cite{Strominger:2016wns} 
 with $G(z) = z^\alpha$, and $\alpha$ a real parameter.  The resulting
 spacetime after acting with this superrotation was argued to contain a cosmic string
 which pierces null infinity at the singular points $z = 0, \infty$.  A major motivation of the present
 paper is to critically examine this proposal, and to further investigate the resulting geometry.  
 
We employ the exact Comp\'ere-Long metric described above, as opposed to the asymptotic expansion at large $r$, since the large $r$ expansion
breaks down when $V$ from equation (\ref{eqn:V}) diverges, which can happen either
when the Schwarzian blows up, or at $z\rightarrow\infty$ where $1/\gamma(z,\bar z)^2$ 
diverges.\footnote{The Schwarzian really should be viewed as a component of a tensor
as in (\ref{eqn:Tzz}), and so it makes sense to ask where the scalar $V$ constructed
from this tensor diverges, as opposed to just where the components themselves diverge. }  
For the $z^\alpha$ superrotation,
the rotational symmetry about the point $z=0$ on the celestial sphere is preserved.  To make this 
symmetry manifest, it is useful to switch to angular coordinates on the sphere, defined by 
$
z = e^{i\phi}\cot\frac\theta2 , \bar z = e^{-i\phi} \cot\frac\theta2,
$
which gives $dz = z(-\frac{d\theta}{\sin\theta} + id\phi)$.  The Schwarzian for the transformation
is $\{z^\alpha,z\} = -\frac{\alpha^2-1}{2 z^2}$, and using the expressions 
(\ref{eqn:gCL}), we obtain the metric components in the $(u,r,\theta,\phi)$
coordinates,
\begin{subequations}
\label{eqn:zalphamet}
\begin{align}
g_{uu}&=-1 - \frac{2u\tau^2}{\rho} \label{eqn:guu} \\
g_{ur} &= -\frac{r}{\rho} \\
g_{u\theta} &= 2 \cot \theta \frac{u^2\tau^2}{\rho} \\
g_{\theta\theta} &= (\rho + u\tau)^2  \label{eqn:gthth}\\
g_{\phi\phi} &= (\rho - u\tau)^2\sin^2\theta, \label{eqn:gphph}
\end{align}
\end{subequations}
where 
\begin{align}
\tau &= \frac{\alpha^2-1}{2\sin^2\theta} \label{eqn:tau} \\
\rho &= \sqrt{r^2+u^2\tau^2}. \label{eqn:rho}
\end{align}
Being locally a diffeomorphism of Minkowski space, the 
Riemann tensor  of this metric vanishes. 
However, there can be curvature defects in regions not covered by  the Bondi coordinate system.  
The coordinates become singular at $\theta = 0,\pi$ due to the blow up in $\tau$, and also at $r=0$, where the spatial metric degenerates.  This allows for the 
possibility of a cosmic string defect in the 
region where the coordinate system degenerates,
 with the string piercing the sphere at the poles.  

To detect whether such defects are present in the above metric, we can compute the 
parallel transport of a vector around the punctures.   If there is a nontrivial contribution 
to the curvature at a cosmic string defect, a vector parallel transported around a loop encircling
the defect will not return to itself.  Since the curvature vanishes in the region covered by the 
$(u,r,\theta,\phi)$ coordinates, the result of the parallel transport will be independent of the 
choice of loop.  Hence, we can choose the loop to follow the $\phi$ coordinate, and use rotational symmetry to simplify the problem.  The parallel transport equation to be solved
is 
\beq\label{eqn:dphV}
\partial_\phi V^\mu = -\Gamma^\mu_{\phi\nu}V^\nu.
\eeq
Since the Christoffel symbol is independent of $\phi$, this equation can easily be exponentiated 
by finding the eigenvalues of $\Gamma^\mu_{\phi\nu}$, viewed as a matrix with indices $\mu,
\nu$.  

The parallel transport computation simplifies significantly after transforming to a different radial coordinate.
Taking $\rho$, defined in equation (\ref{eqn:rho}), as the new radial coordinate, the metric components become 
\begin{subequations}
\label{eqn:gNU}
\begin{align}
    g_{uu} &= g_{u\rho} = -1 \label{eqn:guuNU}\\
    g_{\theta\theta}&= (\rho + u \tau)^2 \\
    g_{\phi\phi} &= (\rho - u \tau)^2 \sin^2\theta, \label{eqn:gphphNU}
\end{align}
\end{subequations}
with all other components vanishing.  This choice of radial coordinate puts the metric into 
Newman-Unti gauge, where $\rho$ is an affine parameter along the radial null geodesics 
\cite{Newman:1962cia, Barnich2012}.   
We will comment further on the relation between the Bondi and Newman-Unti gauge choices 
in  section \ref{section:superrotation}; 
for now, we simply note that the absence of off-diagonal $g_{uA}$ metric components makes
calculating the Christoffel symbols less cumbersome. 

Due to $\phi$-independence of the metric components 
(\ref{eqn:gNU}), the only nonzero  elements in the parallel transport matrix
are $\Gamma^\phi_{\phi \nu}$
and $\Gamma^\mu_{\phi\phi}$, with $\Gamma^\phi_{\phi\phi}=0$.  This means that 
$\Gamma^\mu_{\phi \nu}$ has rank 2 and vanishing trace, so the eigenvalues must be given
by $\lambda_\pm = \pm \sqrt{\frac12\Gamma^\mu_{\phi\nu}\Gamma^\nu_{\phi \mu}} $.
One can then compute
\begin{align}
\frac12\Gamma^\mu_{\phi\nu} \Gamma^\nu_{\phi\mu}
&= -g^{\alpha\beta}\partial_\alpha \sqrt{g_{\phi\phi}} \partial_\beta\sqrt{g_{\phi\phi}}  \nonumber \\
&=  %
(\partial_\rho\sqrt{g_{\phi\phi}})^2-2\partial_\rho\sqrt{g_{\phi\phi}} \partial_u\sqrt{g_{\phi\phi}} 
+\frac{(\partial_\theta\sqrt{g_{\phi\phi}})^2}{g_{\theta\theta}} \nonumber  \\
&= %
\sin^2\theta (1+2\tau) + \frac{\cos^2\theta (\rho + u \tau)^2}{(\rho+u \tau)^2}  \nonumber \\
&= -\alpha^2%
,
\end{align}
yielding eigenvalues $\lambda_\pm = \pm i\alpha$.  
The parallel transport equation
(\ref{eqn:dphV}) can then easily be solved for the eigenvectors $V_{\pm}^\nu$,
\beq
V_\pm^\nu(\phi) = e^{\mp i \alpha \phi} V_{\pm}^\nu(0).
\eeq
The endpoint of a closed loop is at $\phi = 2\pi$, with $V_\pm^\nu(2\pi) = e^{\mp i 2\pi \alpha} 
V_{\pm}^\nu(0)$, and so only returns to itself when $\alpha$ is an integer, signalling 
the presence of curvature in the region interior to where the coordinate
system becomes singular for any $\alpha\neq 1$.\footnote{Even though the vector returns to itself for 
any integral value of $\alpha$, it winds around too many times except when $\alpha = 1$.}  
This result is consistent with the interpretation given in \cite{Strominger:2016wns} that 
the superrotation leads to cosmic string defects in the bulk.

The above conclusion relies crucially on the periodicity of the $\phi$ coordinate being $2\pi$.  
Since the metric is rotationally symmetric in $\phi$, imposing a different periodicity 
condition should have no effect other than to change the conical defect angle.  In particular, 
giving $\phi$ a period of $\frac{2\pi}{\alpha}$  results in the vector returning to itself after 
a single winding, which is consistent with no conical defect in the interior.  Our proposal for
the result of acting with a superrotation is to make such a choice for the $\phi$-periodicity, so
that no cosmic string defects are produced.  

Restricting the range of $\phi$ in this manner has a nontrivial effect on the celestial sphere 
intrinsic geometry.  Since the large $r$ limit of $g_{AB}$ in (\ref{eqn:qAB}) is simply $r^2$ times
the round sphere metric, changing the range of $\phi$ results in a sphere with a conical defect.  
Hence, the price to be paid for having a defect-free bulk metric is to introduce a defect on the sphere at infinity.  
Many definitions of asymptotic flatness require the spatial metric to approach a sphere with 
no defects as $r\rightarrow\infty$.  This means the superrotated spacetimes with no bulk conical defects 
are asymptotically flat only in a weaker sense, in which more general metrics on the celestial sphere are allowed.  
We will further justify this prescription to restrict the range of $\phi$ in 
section \ref{sec:funddomain}, after discussing the transformation in detail.

\section{Geometry of spatial surfaces}\label{sec::geometry}

The possibility for conical defects in the Comp\`ere-Long metric (\ref{eqn:gCL}) 
is not its only nonintuitive 
feature.  Another crucial aspect is that the large $r$ expansion is only valid in regions
where $V$ is bounded.    
Near divergences in $V$, the spatial metric differs significantly
from the sphere metric.  There is an important order of limits between  $r$ going to infinity and approaching singularities
in $V$.  Taking $r$ large first results in a picture of a celestial sphere with punctures at the points where $V$ diverges.  In this section we discuss the opposite order of limits, keeping $r$ finite and 
exploring the geometry of constant $(u,r)$ surfaces near the puncture points.  We also analyze the behavior of the spatial
surfaces at constant $(u,\rho)$ in Newman-Unti gauge, which share some features with the Bondi gauge spatial surfaces, but 
differ substantially near the punctures.  

\subsection{Bondi surfaces} \label{sec:bondisurf}
The induced metric on surfaces of constant $(u,r)$ is obtained from equations (\ref{eqn:gthth}), (\ref{eqn:gphph}),
\beq \label{eqn:bondispheremetric}
ds^2 = \left(u \tau + \sqrt{r^2 + u^2\tau^2}\right)^2 d\theta^2 + \sin^2\theta \left(u \tau - \sqrt{r^2+u^2\tau^2}\right)^2 d\phi^2
\eeq
with $\tau$ given by (\ref{eqn:tau}).   This metric is symmetric under the reflection $\theta\rightarrow \pi-\theta$, 
and the divergence occurs in $\tau$ at $\theta = 0,\pi$.  Away from these points, the metric approaches
the round sphere metric $r^2(d\theta^2 + \sin^2 \theta d\phi^2)$ when $r\gg |u\tau|$.  However, near the poles where $\tau$ diverges,
the opposite limit $|u\tau| \gg r$ must be considered.  In this case, two different limiting metrics are obtained, depending on the sign 
of $u\tau$,
\beq \label{eqn:limitgeom}
ds^2 = 
\begin{cases}
4 u^2 \tau^2 d\theta^2 + \frac{r^4}{4u^2\tau^2} \sin^2\theta d\phi^2 & u\tau>0 \\
\frac{r^4}{4 u^2 \tau^2} d\theta^2 + 4 u^2 \tau^2 \sin^2\theta d\phi^2 & u\tau < 0
\end{cases}
\eeq

We begin by examining the qualitative features of the metric with $u\tau>0$, which, for $\alpha>1$, corresponds to the 
region to the future of the $u=0$ lightcone.
Near $\theta = 0$, the $d\theta^2$ coefficient approaches $\frac{u^2(\alpha^2-1)^2}{\theta^4}$, which implies the proper length from 
a finite value of $\theta$ to the pole diverges, 
$\displaystyle u(\alpha^2-1)\int_{\theta_0}^0 \frac{d\theta}{\theta^2} \rightarrow \infty $.  This is drastically different
from the sphere metric, for which the distance to the pole is finite.  At small values of $\theta$, the $d\phi^2$ coefficient 
goes like $\frac{4r^4u^2}{(\alpha^2-1)^2}\theta^6$, indicating that the length of the circles at constant $\theta$ are shrinking 
proportional to $\theta^3$ at small values of $\theta$.  

The picture this suggests for the constant $(u,r)$ surface is one which is roughly spherical near the equator $\theta=\pi/2$, but which
becomes narrow and pointed near the poles.  We call these oblong surfaces ``cosmic footballs'' due to their resemblance to an 
(American) football, although the tips of the cosmic footballs extend to an infinite distance.  Note that this infinite distance
to the pole can only be possible if the surface touches null infinity at $\theta=0$.  Hence, we can conclude that the singular
point of the superrotation not only introduces punctures into the celestial sphere, but in fact maps a whole plane of points 
at $\theta=0$ out to infinity, a fact that seems to have been underappreciated previously.  At small values of $\theta$, the geometry
resembles a thin tube, which encloses the string defect if it is present.  The surface runs off to infinity 
rather than allowing the string to intersect it.  

\begin{figure}[t]
    \centering
    \includegraphics[width=\linewidth]{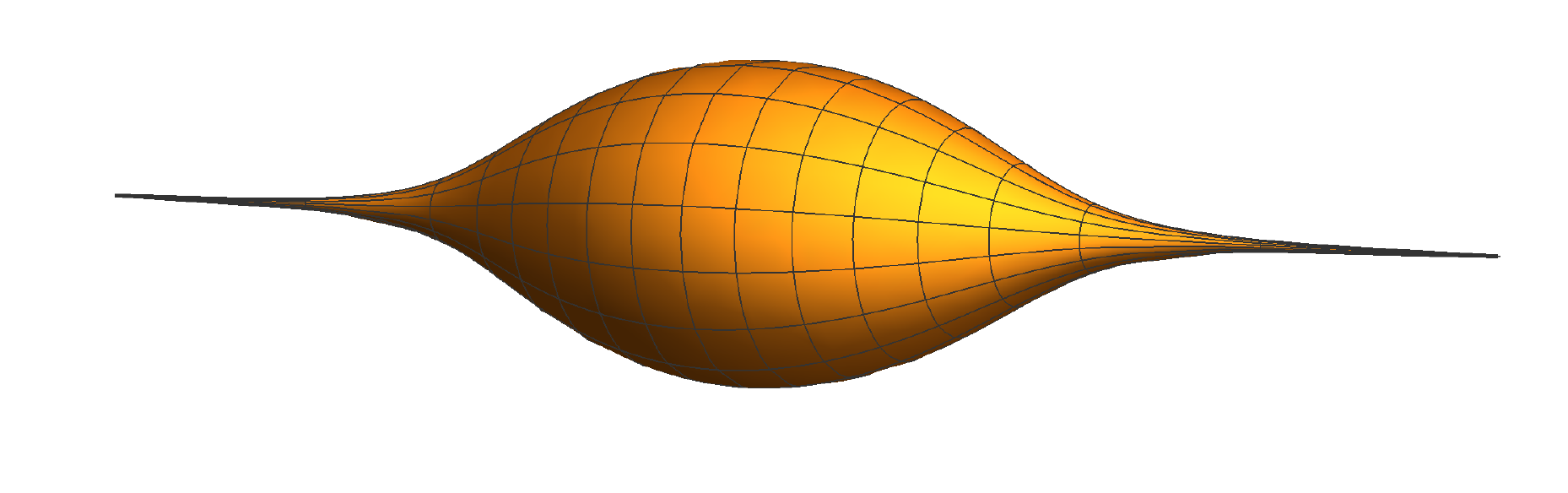}
    \caption{Isometric embedding of the cosmic football $(u=1,r=4,\alpha=1.4)$ into $\mathbb{R}^3$ 
    }
    \label{fig::embedR3pos}
\end{figure}

The cosmic football intrinsic geometry can be better visualized by embedding the geometry into 
$\mathbb{R}^3$. 
Since the surface has $\phi$-rotational symmetry, we employ cylindrical coordinates 
$(R, \Phi, Z)$ on $\mathbb{R}^3$.  The $\Phi$ coordinate has periodicity of $2\pi$, and so for surfaces
for which the cosmic string defect has been eliminated by adjusting the periodicity of its intrinsic
$\phi$ coordinate, we must set $\Phi = \alpha \phi$.  The embedded surface is 
then defined by finding $Z(\theta)$ and $R(\theta)$ such that the induced metric matches
(\ref{eqn:bondispheremetric}).  This leads to the condition
\beq
\Big(Z'(\theta)^2+R'(\theta)^2\Big) d\theta^2 + R(\theta)^2 \alpha^2 d\phi^2
= (u\tau + \rho)^2 d\theta^2 + (u\tau - \rho)^2 \sin^2\theta d\phi^2.  
\eeq
Then $R(\theta)=(u\tau-\rho)\sin\theta$, after which $Z(\theta)$ satisfies the 
differential equation
\beq
Z'(\theta) = \sqrt{(u\tau + \rho)^2 - R'(\theta)^2},
\eeq
which can be numerically integrated to produce plots of the embedded surface.  An 
example of such a surface with $u\tau>0$ is depicted in figure \ref{fig::embedR3pos}.  
The result is the cosmic football surface described above, spherical near the equator $\theta = \pi/2$,
and infinitely long and pointed near the poles.  

The second case to consider are the geometries with $u\tau<0$, which lie at $u<0$ when $\alpha>1$.  These surfaces have even stranger
properties than the cosmic footballs described above.  In the small $\theta$ limit, (\ref{eqn:limitgeom}) indicates that
the length from a finite value of $\theta$ to the pole
is now finite, $\displaystyle \frac{r^2}{u(\alpha^2-1)} \int_{0}^{\theta_0} \theta^2 d\theta \rightarrow \text{finite} $, 
which appears to be consistent with a metric that is topologically a sphere.  However, the lengths
 of the constant $\theta$ circles
diverge as $\theta^{-1}$, indicating that the geometry flares out near the pole.  This divergence in the circle circumference
within a finite proper distance along $\theta$ suggests the development of a curvature singularity, and one can verify that the 
scalar curvature of the surface is proportional to $\theta^{-6}$ as $\theta\rightarrow 0$.  
The flaring behavior of the surface indicates a region of negative curvature near the pole, and we refer to these surfaces as 
``cosmic horns'' since the geometry resembles the bell of a horn.  

\begin{figure}[t]
    \centering
    \includegraphics[width=0.3\textwidth]{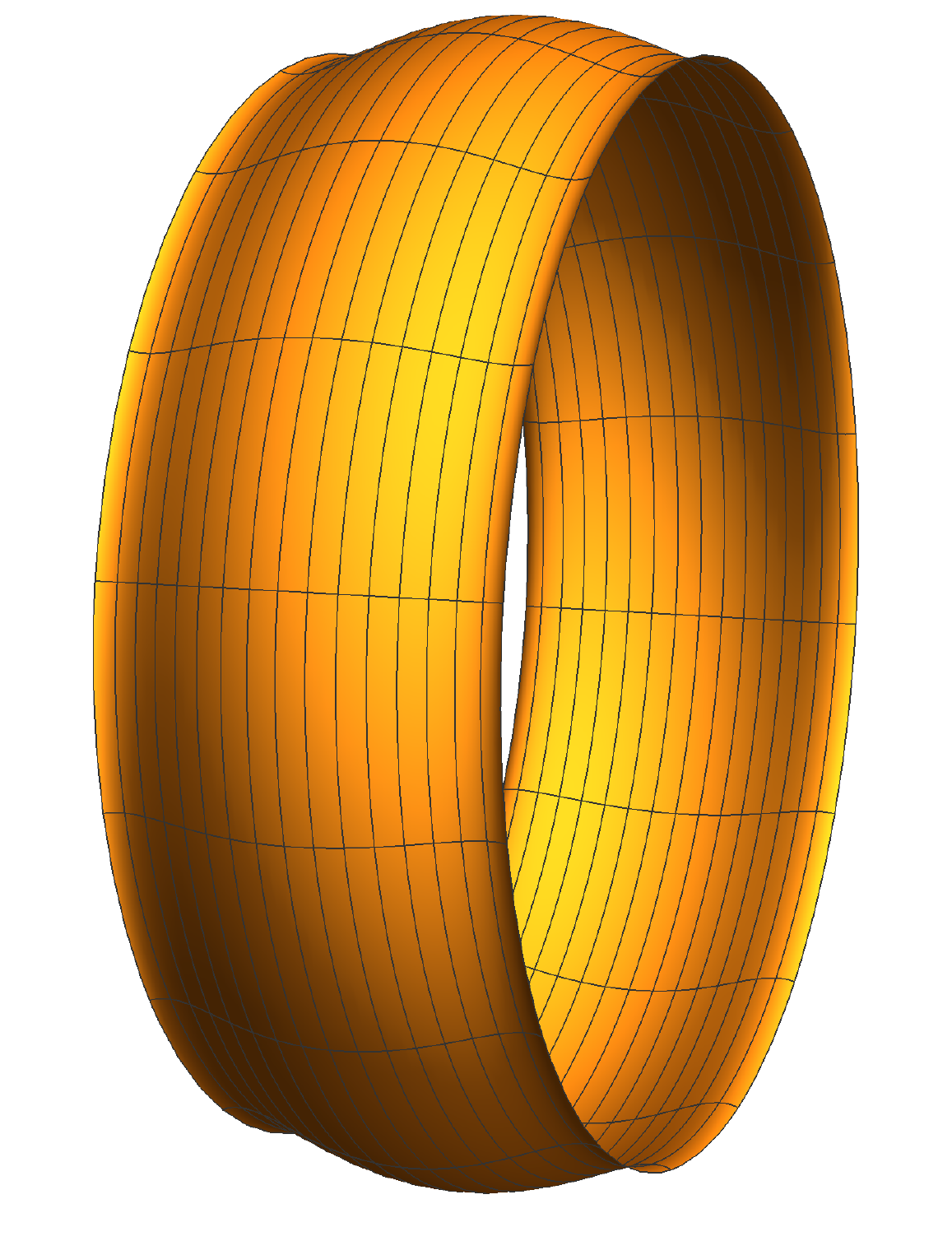}
    \caption{Isometric embedding of the cosmic horn $(u=-1,r=0.6,\alpha=0.8)$ into $\mathbb{R}^3$.  
    The embedding breaks down shortly after it begins to flare out,
    becoming too negatively curved to embed into $\mathbb{R}^3$ in a way that preserves
    rotational symmetry.
    }
    \label{fig::embedR3neg}
\end{figure}

Embeddings of the cosmic horn geometry into $\mathbb{R}^3$ can be produced in a similar manner to the cosmic footballs. 
In this case, however, a rotationally symmetric embedding exists for only some finite 
range of $\theta$ near $\pi/2$.  At a certain critical
value, the surface becomes too negatively curved to embed nicely into Euclidean space,
as can be seen in figure \ref{fig::embedR3neg}.  Although the nonexistence of 
a full embedding sounds counterintuitive, it is worth recalling
that this geometry naturally arose as a codimension-two surface embedded in four dimensional \emph{Lorentzian} space, which
can accommodate negatively curved surfaces more easily.  The embedding of these surfaces into Minkowski space are discussed in
more detail in section \ref{section:superrotation}, 
however, we can intuit their general features from the above discussion.  The diverging length
of the constant $\theta$ circles suggest that the surface must again intersect null infinity at $\theta=0$, just as the $u\tau>0$ 
surfaces did.  In order to reach null infinity along a curve with finite proper length, it must be that the surface is rapidly 
becoming null near the pole.  

\subsection{Newman-Unti surfaces} 

Due to the relative simplicity of the Newman-Unti form of the metric, we also examine the geometry of the 
spatial surfaces in Newman-Unti gauge, which are the level sets of $u$ and $\rho = \sqrt{r^2+u^2\tau^2}$.  
These surfaces display variegated behaviors depending on the values of $\rho$, $u$, and $\alpha$, 
some of which are explored
in section \ref{section:superrotation}.  
Here we discuss their general properties.  The induced metric on a
Newman-Unti surface  is
\beq \label{eqn:NUmetric}
ds^2 = (\rho + u \tau)^2 d\theta^2 + (\rho-u\tau)^2 \sin^2\theta d\phi^2.
\eeq
The original Bondi coordinates cover the region $r>0$, which translates to the region $\rho>|u\tau|$ in Newman-Unti coordinates.  
Bondi coordinates are designed to break down at $r=0$, which coincides to the location where the null congruence tangent to 
$\partial_r^a$ forms caustics.  Since Newman-Unti coordinates are constructed from the same congruence and differ 
only in the choice of parameter along the radial generators, 
we should expect to find the same singularities associated with 
caustics when $\rho\rightarrow |u\tau|$.  This singularity is apparent in the induced metric (\ref{eqn:NUmetric}), where 
one or the other components vanishes in this limit, depending on the sign of $u\tau$.  This is actually a curvature singularity, 
since the scalar curvature is given by, 
\beq
\mathcal{R}_\text{NU} = \frac{2}{\rho^2-u^2\tau^2}.
\eeq

Since $|\tau|\rightarrow\infty$ near $\theta=0$, for any fixed value of $\rho$, there will always be a region near the poles for
which $\rho<|u\tau|$, and hence goes beyond the region covered by the Bondi coordinates.  
Starting with $\rho>\left|\frac{\alpha^2-1}{2} u\right|$, the region near $\theta = \pi/2$ will lie in the Bondi patch. As 
$\theta$ decreases to move toward the pole, a critical value will be reached where $|u\tau| = \rho$ and the surface becomes
singular as it exits the Bondi patch.  No other singularities develop as $\theta$ is decreased further to $0$.  As was done for the 
Bondi spheres, the distance to the pole can be computed, and it is seen to diverge as $\displaystyle
\left|u\frac{\alpha^2-1}{2} \right|\int_0^{\theta_0} \frac{d\theta}{\theta^2}\rightarrow\infty$, again implying that $\theta=0$ must 
reside at infinity. Where exactly it intersects null infinity requires the more detailed analysis of 
section \ref{section:superrotation}, but 
we again are able to conclude that the large $\rho$ expansion is not the same as an expansion near infinity, since surfaces at 
any value of $\rho$ intersect infinity.

\section{General structure of finite superrotations }
\label{section:superrotation}

In the previous section, we saw that surfaces at finite radius and $u$ tended to extend out to infinity,
both for Bondi and Newman-Unti gauges.  To better understand 
where exactly these surfaces intersect null infinity and how they are situated inside of 
Minkowski space, it is helpful to work out the explicit expression for the finite superrotation 
transformation.  Comp\`ere and Long first derived this transformation in
\cite{Compere2016b}, and used it to write the 
form of the pulled back metric in \cite{Compere:2016jwb}.  
Here we give an alternative derivation of the finite superrotation,
which has the advantage of making it clear precisely how points in Minkowski space are
mapped around.  

The details of the derivation are given in appendix \ref{sec::diffeo}, but the idea behind the transformation can be summarized here.
We begin with Minkowski space parameterized in retarded radial coordinates $(u,r, z, \bar z)$, in which the metric takes the form
\beq
ds^2 = -du^2 -2 du dr + 2 r^2 \gamma(z,\bar z) dz d\bar z,
\eeq
with $\gamma(z,\bar z)$ as in equation (\ref{eqn:gzbarz}).  The superrotation transformation can be specified by 
giving the image $(t,\vec{x})$ in Cartesian coordinates of each original point of Minkowski space, i.e.\ by determining
$t$ and $\vec{x}$ as functions of $(u,r,z,\bar z)$.  Once this is done, the superrotated metric is obtained by pulling 
back the standard Minkowski metric by this transformation.  A superrotation is defined to act on the celestial sphere 
$u=0, r\rightarrow\infty$ as a holomorphic transformation, $z \mapsto G(z)$, and its action elsewhere can be fixed 
by demanding that the pulled back metric components satisfy the Bondi gauge conditions.  Near $\scri^+$, the infinitesimal vector
field generating the transformation is independent of the metric components, and so the flow can be solved at the first few orders
in the large $r$ expansion by integrating the flow of this vector field.  Beyond these leading orders, the vector field 
generating the flow is dependent on components of the metric, and hence changes as one flows along it.  However, once 
the flow is integrated near $\scri^+$, it can be solved at a finite distance into the bulk by simply imposing the Bondi gauge
conditions.  

Bondi gauge requires that the radial vector $\partial_r^a$ be tangent to null geodesics.  Each such null geodesic intersects
$\scri^+$ at a single location, and the asymptotic solution determines which null geodesic the image point lies on.  The points 
on a null geodesic with parameter $t$ have Cartesian coordinates $(t,\vec{a} + t \hat \omega)$, 
with the geodesic completely specified by the
the fixed unit vector 
$\hat \omega$ and fixed vector $\vec a$.  This means that the superrotation transformation
takes the form
\begin{subequations}
\label{eqn:nullgeo}
\begin{align}
x^0 &= t \label{eqn:x01}\\
\vec x &= \vec b + c \hat \omega + t \hat\omega, \label{eqn:xvec}
\end{align}
\end{subequations}
with $\vec a = \vec b + c \hat\omega$,  $\vec b \cdot \hat \omega = 0$, and 
all objects are functions of the initial coordinate $(u,r,z,\bar z)$, but only $t$ can depend on $r$, since $r$ must be 
a parameter along the geodesic at fixed $(u,z,\bar z)$.  The asymptotic solution to the flow determines $\vec b$, $c$ and 
$\hat\omega$ in terms of $(u, z, \bar z)$, and then $t$ is fixed by imposing the final Bondi gauge condition, that 
the pulled back spatial metric on a constant $(u,r)$ surface has determinant $-r^4\gamma(z,\bar z)^2$.  When all is said and 
done, this results in the following functions describing the $G(z)$ superrotation:
\begin{subequations}
\label{eqn:finitesuperrot}
\begin{align}
t(u,r,z,\bar z) &= \frac{1}{\Omega} \left(u L + \sqrt{r^2 + u^2 V}\,\right) \\
\Omega(z,\bar z) &= \left[ \partial G \bar \partial \bar G \frac{\gamma(G, \bar G)}{\gamma(z,\bar z)}
\right]^{1/2} 
\label{eqn:Omega} \\
\gamma(z, \bar z) &= \frac{2}{(1+ z\bar z)^2} \\
L(z,\bar z) & = \frac{1}{\gamma(z,\bar z) } \partial \log \Omega\, \bar \partial\log\Omega 
+ \frac12(\Omega^2 +1) \\
V(z,\bar z) &= \frac{1}{4 \gamma(z,\bar z)^2} \{G(z),z\} \{\bar G(\bar z), z\} \\
c(u,z,\bar z) & = - u \Omega \\
\hat\omega(z,\bar z) &= \frac{1}{1+G\bar G} 
\begin{pmatrix} G+\bar G \\ -i(G-\bar G) \\ -1+G\bar G\end{pmatrix}\\
\vec b(u, z,\bar z) & = \frac{u}{\Omega}\left[\frac{\partial \hat\omega}{\partial G} H 
+ \frac{\partial\hat\omega}{\partial\bar G} \bar H\right]
= \frac{u}{\Omega(1+G\bar G) } \left[(-H \bar G -\bar H G)\hat \omega
 + \begin{pmatrix} H + \bar H \\ -i(H-\bar H) \\ H \bar G + \bar H G\end{pmatrix} \right] \\
H( z,\bar z) & = -\frac{\partial G}{\gamma(z,\bar z)} \bar \partial \log \Omega \\
\bar H( z,\bar z) &= -\frac{\bar \partial\bar G}{\gamma(z,\bar z) } \partial\log\Omega
\label{eqn:Hbar}
\end{align}
\end{subequations}

A similar procedure can be used to derive a superrotation that instead preserves 
Newman-Unti gauge.\footnote{Recall that the asymptotic symmetries are defined as 
the group of all transformations that act on the spacetime and its boundary, modulo
those that fix the boundary \cite{Wald1984}.  
This means that the extension of the symmetry into the bulk
has a large degree of arbitrariness, which is why preserving Newman-Unti over Bondi 
gauge results in a different bulk transformation which nevertheless represents the same 
asymptotic symmetry.
Here, we distinguish the transformation by 
denoting the radial coordinate $\rho$ for the Newman-Unti transformation, which also
allows us to relate it to the Bondi radial coordinate $r$.}
This gauge still requires radial vectors to be tangent to null geodesics, and its only difference from Bondi gauge is how the radial coordinate $\rho$ is fixed.  
The Newman-Unti gauge condition requires $\rho$ to be an affine parameter along
the geodesic.  Since $t$ itself is an affine parameter for the null ray at fixed $(u,z,\bar z)$ described by
(\ref{eqn:nullgeo}), it must be related to $\rho$ by an affine transformation, $t = f(u,z,\bar z) + g(u,z,\bar z) \rho$.  
Demanding that $\partial_\rho^\alpha = g^{\alpha\beta}\nabla_\beta u$ fixes $g(u,z,\bar z) = \frac{1}{\Omega}$.  The final condition
is that the $\mathcal{O}(\rho^{-2})$ term in $\partial_\rho \log \det(g_{AB})$ vanish 
\cite{Newman:1962cia, Barnich2012}, 
and this sets $f(u,z,\bar z) = \frac{u L}{\Omega}$.  Hence, the transformation preserving 
Newman-Unti gauge is 
\beq \label{eqn:tNU}
t = \frac{1}{\Omega} (uL + \rho), 
\eeq
and all other quantities are determined exactly as in the Bondi transformation (\ref{eqn:finitesuperrot}).  
It is easy to see from this equation that the Newman-Unti radial coordinate is related to the Bondi radial coordinate
by $\rho = \sqrt{r^2+ u^2 V}$, as mentioned in section \ref{sec:defect}.  

\subsection{Principal domain}
\label{sec:funddomain}
With the explicit formula (\ref{eqn:finitesuperrot}) in hand, we can now justify the prescription described
in section \ref{sec:defect} to eliminate the cosmic string singularity in the $G(z)=z^\alpha$ class of superrotated spacetimes.  
The key point is that the transformation is not a diffeomorphism of Minkowski space; rather, the map
can be many-to-one for some points when $\alpha>1$, or not cover all of Minkowski space when $\alpha<1$.  Adjusting the 
periodicity of the $\phi$ coordinate corresponds to choosing the coordinate range so that the 
image of the map covers Minkowski space exactly once, at least in the region for which the Bondi coordinates are regular.  We refer to such a choice of coordinate range as the principal domain
of the map.  
The fact that the $z^\alpha$ superrotation preserves rotational symmetry makes it easy to find such a principal domain for the transformation.  
The final azimuthal angle $\phi_f$ of the image can be computed in terms of 
the components of $\vec x$ in (\ref{eqn:xvec}) via $\tan \phi_f
= \frac{x^2}{x^1}$.  Recalling the relation $z = \cot\frac\theta2 e^{i\phi}$, 
it is straightforward to show that that $\hat\omega$ and $\vec b$ can be written
\beq
\hat \omega = A(\theta)\begin{pmatrix} \cos \alpha\phi \\ \sin\alpha\phi\\ B(\theta) \end{pmatrix},\qquad
\vec{b} = C(\theta) \begin{pmatrix} \cos\alpha\phi \\ \sin\alpha\phi \\ D(\theta) \end{pmatrix}
\eeq
for some $\phi$-independent functions $A(\theta)$, $B(\theta)$, $C(\theta)$, and $D(\theta)$ whose precise form will not be
needed.  This implies that $\tan \phi_f = \tan \alpha\phi$, or simply $\phi_f = \alpha \phi$.  
Taking the initial $\phi$ to have range $(-\pi,\pi)$, we see that the range of the final angle is $\phi_f\in (-\alpha \pi, \alpha\pi)$.  
Since this exceeds a period of $2\pi$ when $\alpha>1$, we find that multiple initial points can map to the same final point; e.g.,\ 
$(u,r,\theta, \phi)$
and $(u,r,\theta, \phi-\frac{2\pi}{\alpha})$ for $\phi>\frac{\pi}{\alpha}$ map to the same point.  Restricting $\phi$ to lie 
in the principal domain $(-\frac{\pi}{\alpha},\frac\pi\alpha)$ eliminates multivaluedness of the map, and this is the appropriate
range for $\phi$ in order to eliminate the conical defect.  Similarly, for $\alpha<1$, the image of the transformation
only covers a wedge of Minkowski space.  To cover the remainder of Minkowski space, the initial $\phi$ coordinate range must 
be greater than $2\pi$, meaning that the principal domain is a branched cover of Minkowski space.

\subsection{Embeddings into Minkowski space}
\label{sec:embeddings}

%%%%%%
\begin{figure}[t]
    \centering
    \savebox{\largestimage}{\includegraphics[width=0.3\textwidth]{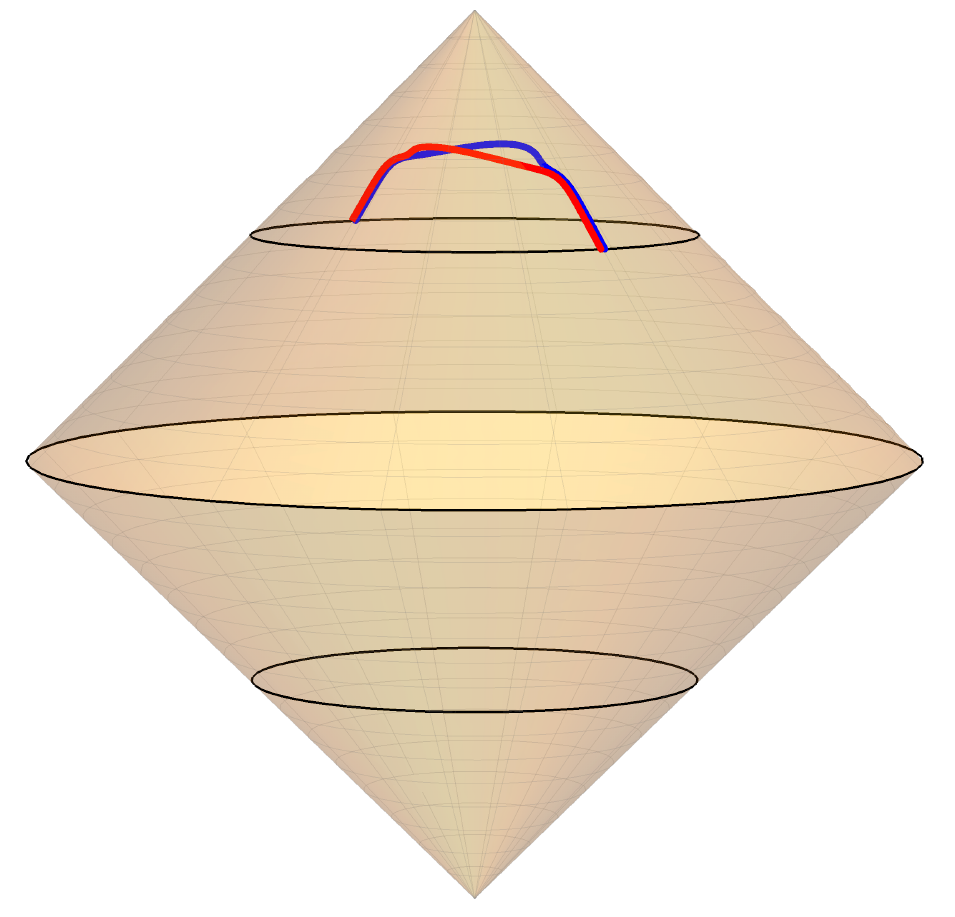}}
    \begin{subfigure}[b]{0.3\textwidth}
        \centering
        \usebox{\largestimage}
        \caption{Bondi surface}
        \label{fig:B1P}
    \end{subfigure}
    \begin{subfigure}[b]{0.3\textwidth}
        \centering
        \raisebox{\dimexpr0.5\ht\largestimage-0.5\height}{
            \includegraphics[width=0.8\textwidth]{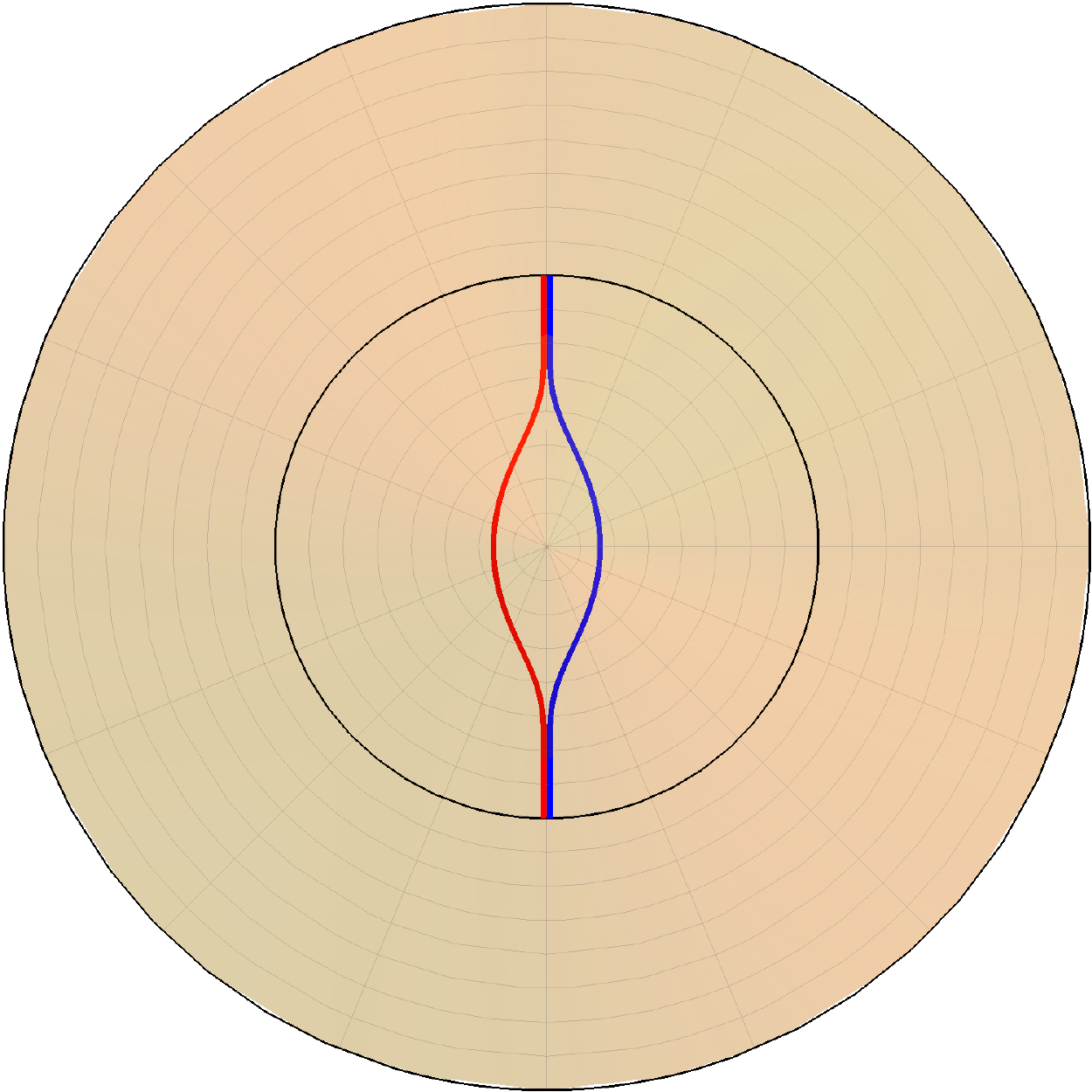}
        }
        \caption{Top view}
        \label{fig:B1T}
    \end{subfigure}
    \hspace{-0.2cm}
    \begin{subfigure}[b]{0.3\textwidth}
        \centering
        \raisebox{\dimexpr0.5\ht\largestimage-0.5\height}{
        		\includegraphics[width=0.9\textwidth]{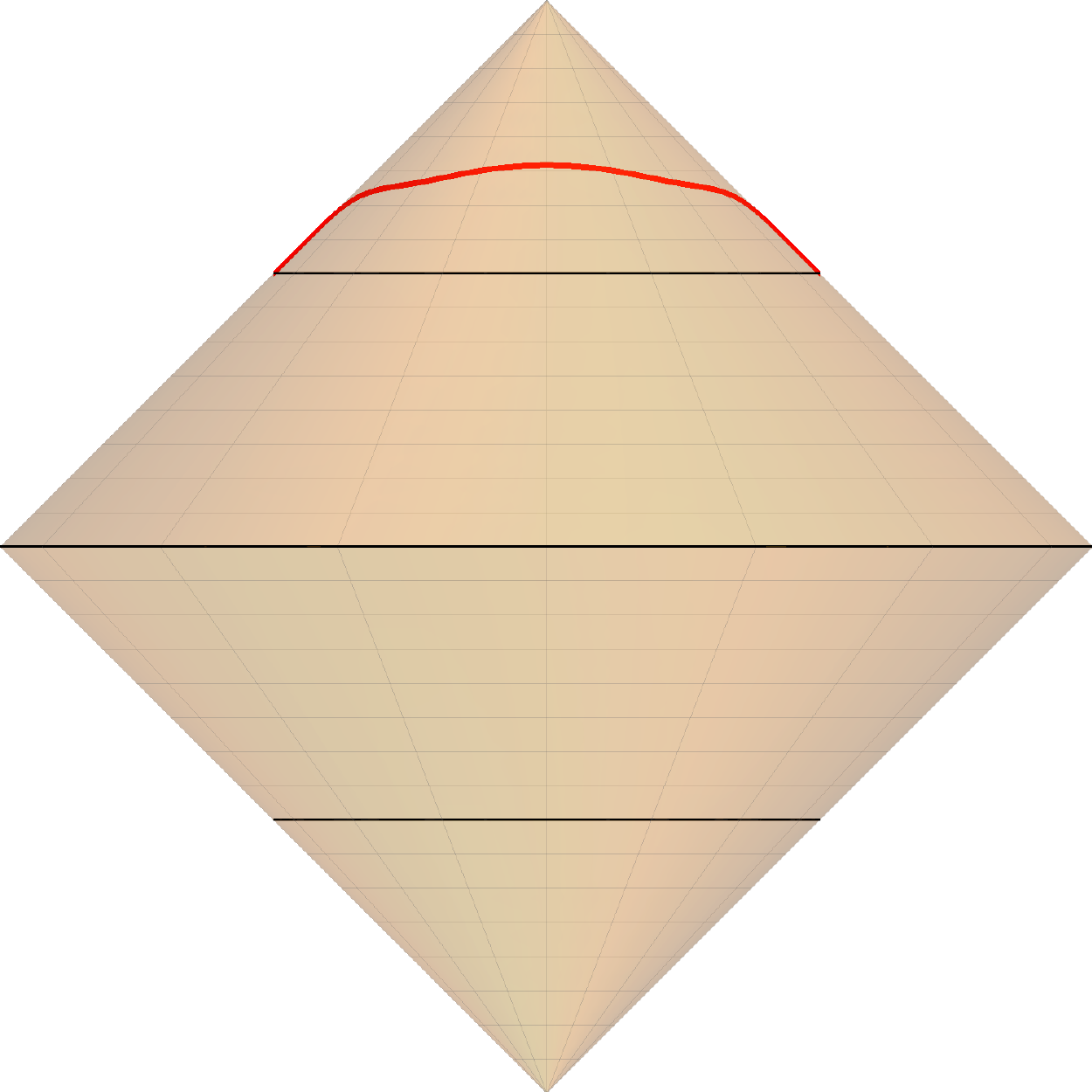}
	}
        \caption{Side view}
        \label{fig:B1X}
    \end{subfigure}
    %%%%%%
    \\
    \savebox{\largestimage}{\includegraphics[width=0.3\textwidth]{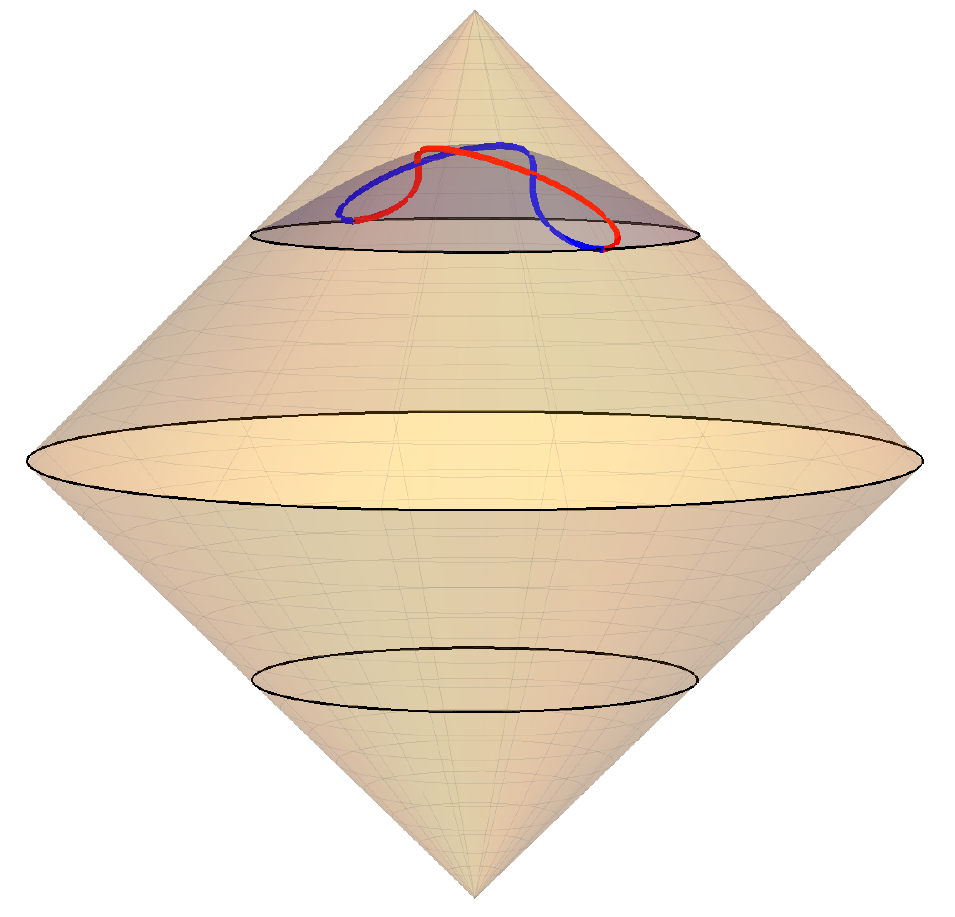}}
    \begin{subfigure}[t]{0.3\textwidth}
        \centering
        \usebox{\largestimage}
        \caption{Newman-Unti surface}
        \label{fig:NU1P}
    \end{subfigure}
    \begin{subfigure}[t]{0.3\textwidth}
        \centering
        \raisebox{\dimexpr0.5\ht\largestimage-0.5\height}{
            \includegraphics[width=0.8\textwidth]{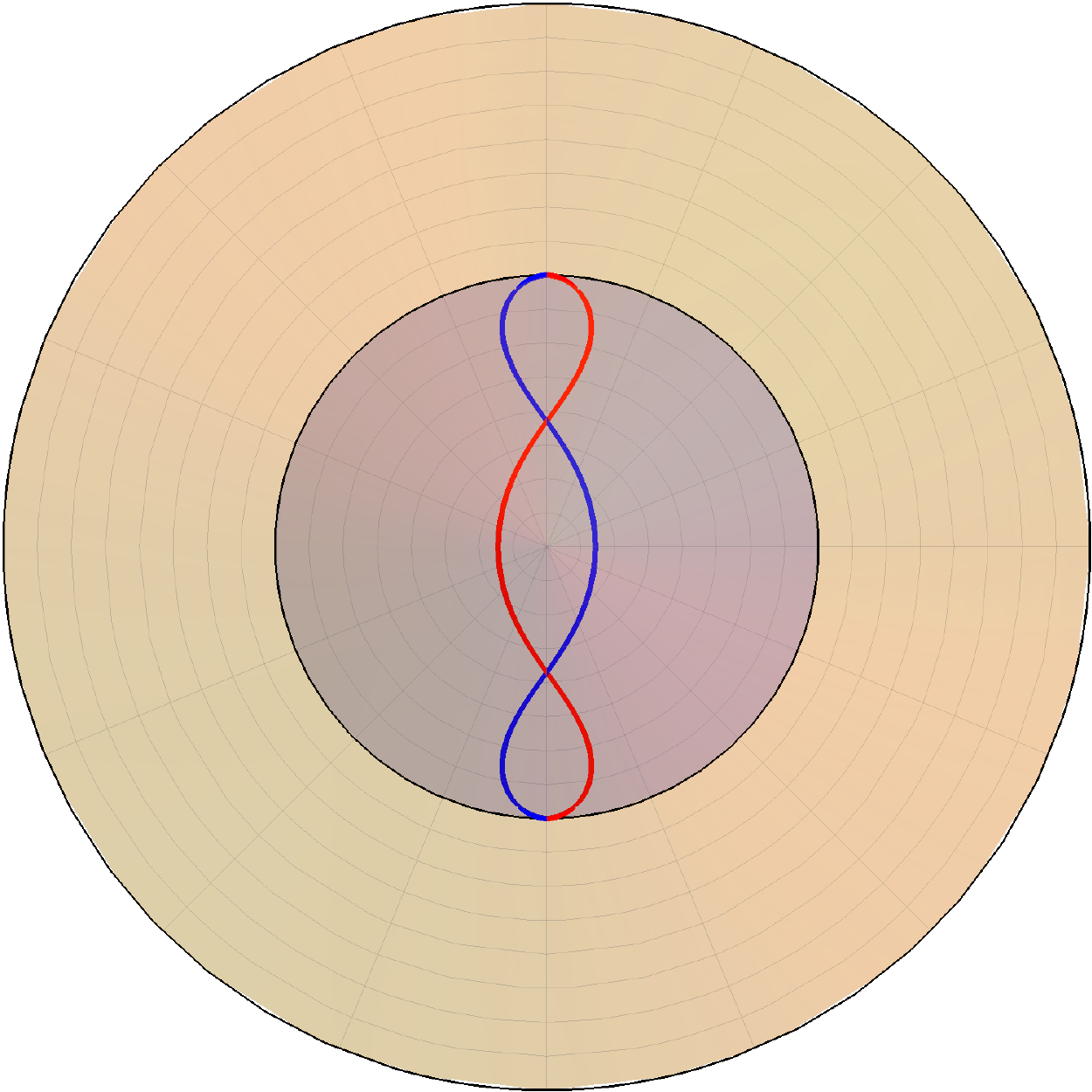}
        }
        \caption{Top view}
        \label{fig:NU1T}
    \end{subfigure}
%    \\
    \hspace{-0.2cm}
    \begin{subfigure}[t]{0.3\textwidth}
        \centering
        \includegraphics[width=0.9\textwidth]{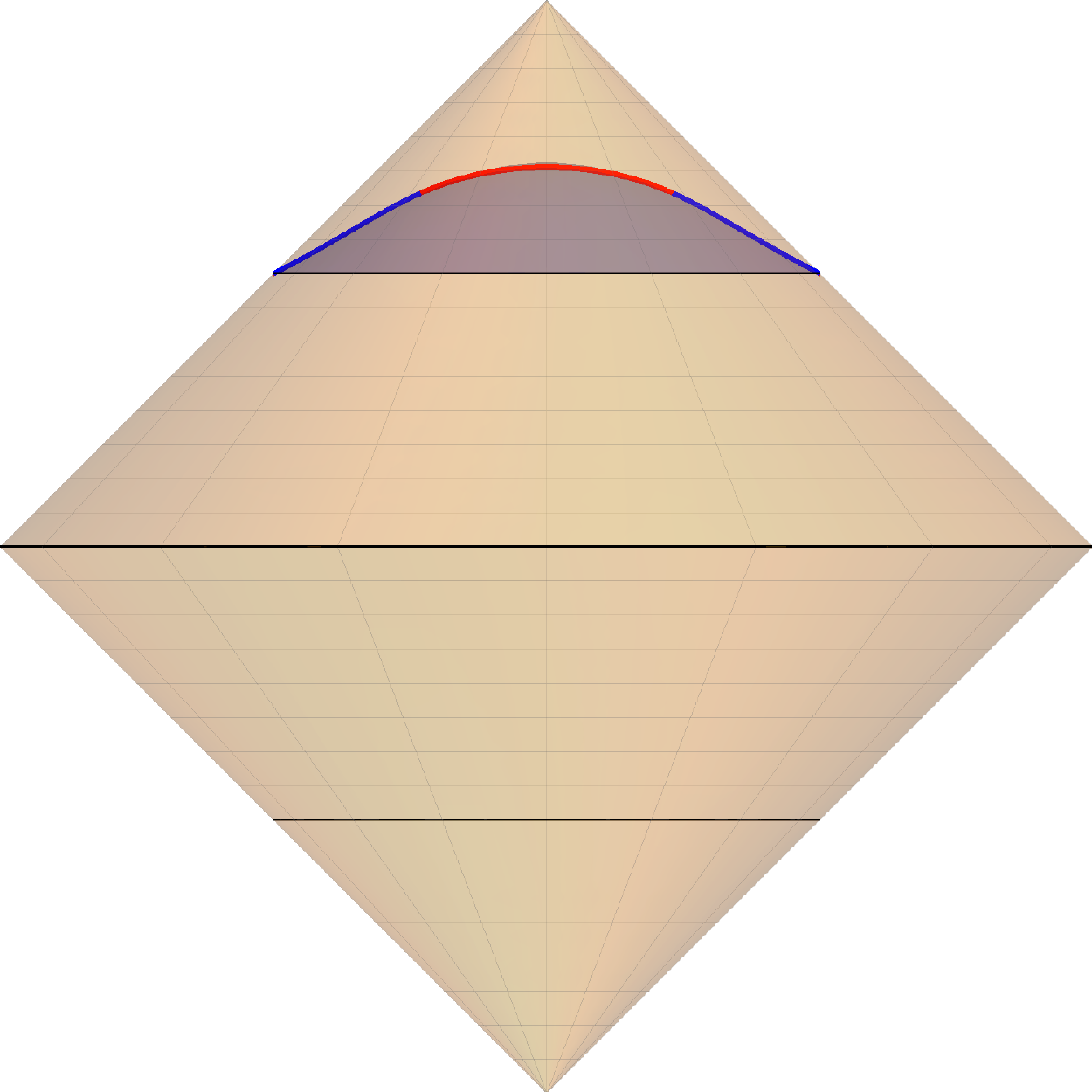}
        \caption{Side view}
        \label{fig:NU1X}
    \end{subfigure}
    \caption{
    %$\alpha = 1.4$, $r=1.48$, $u=1$  
   Bondi and Newman-Unti surfaces with $\alpha>1$, $u>0$.  
   Since the surfaces are rotationally  symmetric, the $\phi$ coordinate is suppressed, and the red and blue curves denote antipodal
   points on the surface at $\phi=0$ and $\phi=\frac\pi\alpha$.  
    }
    \label{fig:a+u+}
\end{figure}

\begin{figure}[t]
    \centering
    \savebox{\largestimage}{\includegraphics[width=0.3\textwidth]{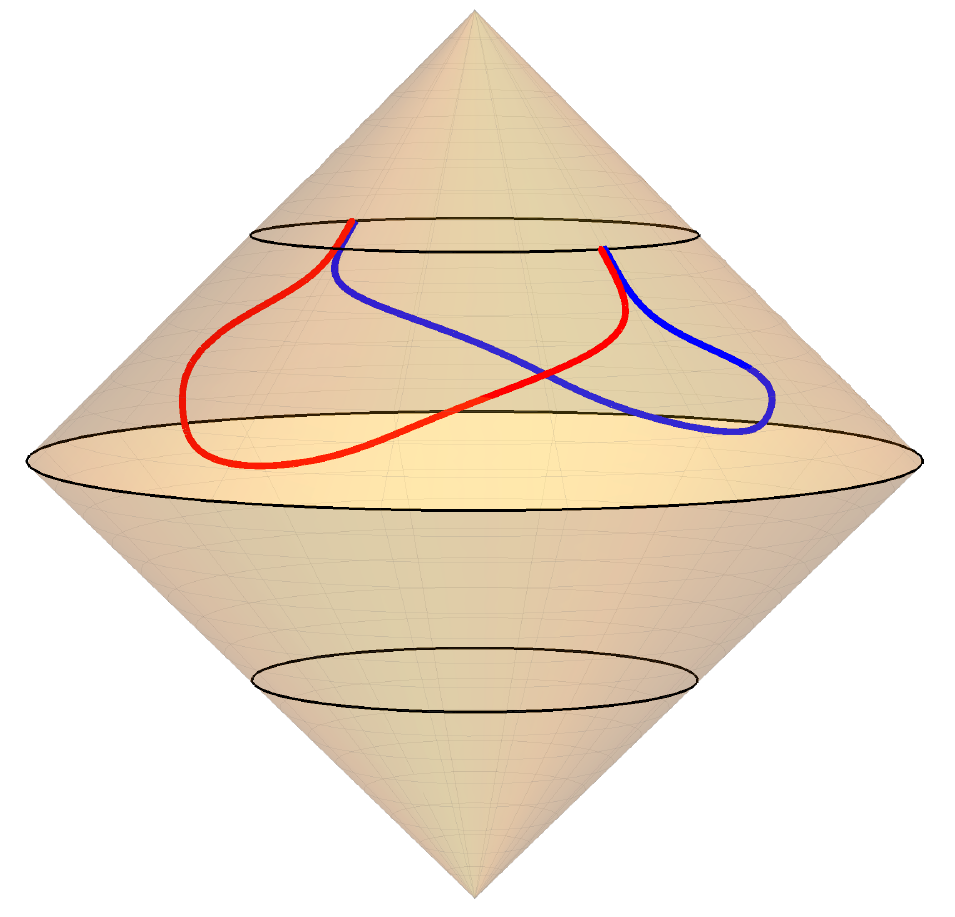}}
    \begin{subfigure}[b]{0.3\textwidth}
        \centering
        \usebox{\largestimage}
        \caption{Bondi surface}
        \label{fig:B2P}
    \end{subfigure}
    \begin{subfigure}[b]{0.3\textwidth}
        \centering
        \raisebox{\dimexpr0.5\ht\largestimage-0.5\height}{
            \includegraphics[width=0.8\textwidth]{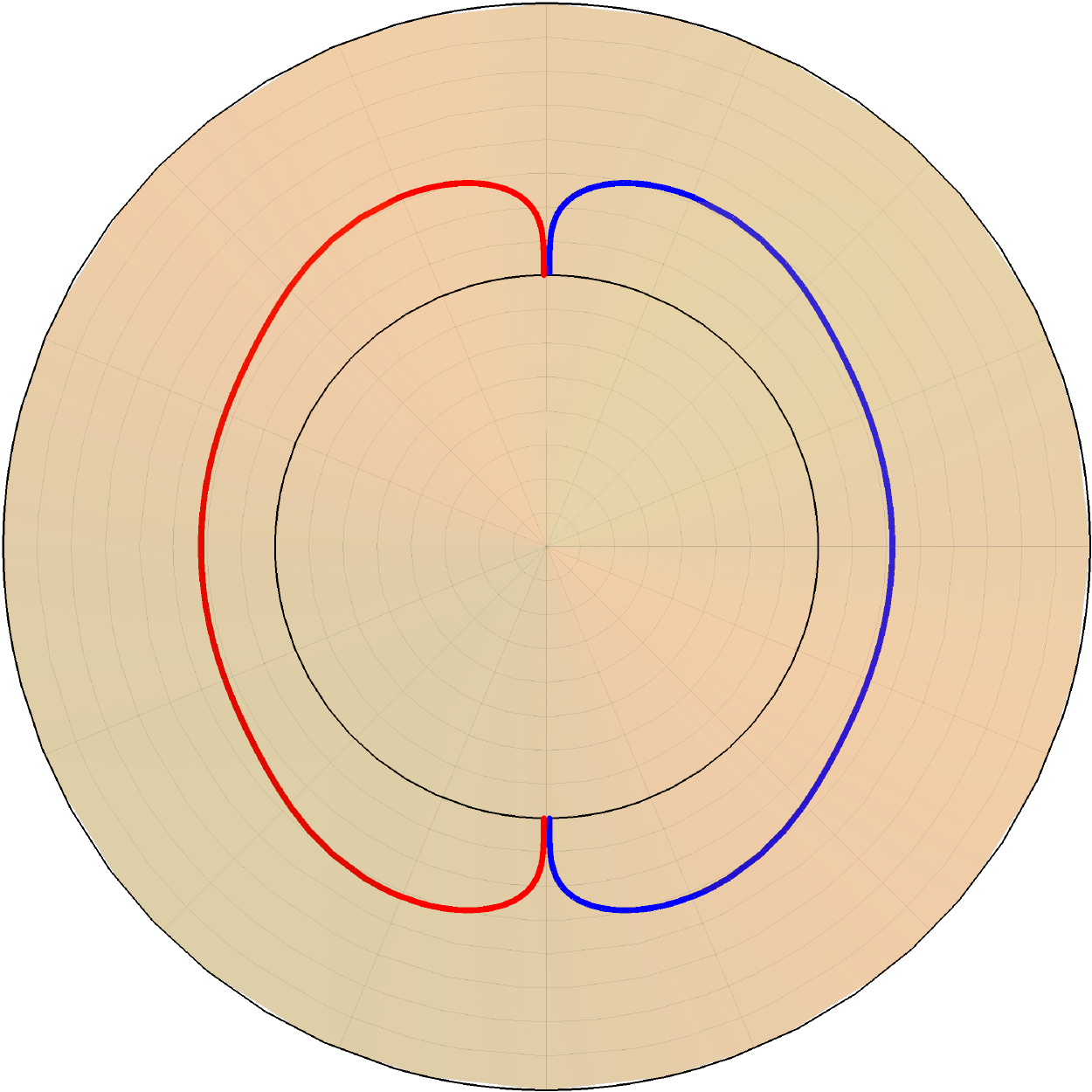}
        }
        \caption{Top view}
        \label{fig:B2T}
    \end{subfigure}
    \hspace{-0.2cm}
    \begin{subfigure}[b]{0.3\textwidth}
        \centering
        \raisebox{\dimexpr0.5\ht\largestimage-0.5\height}{
        		\includegraphics[width=0.9\textwidth]{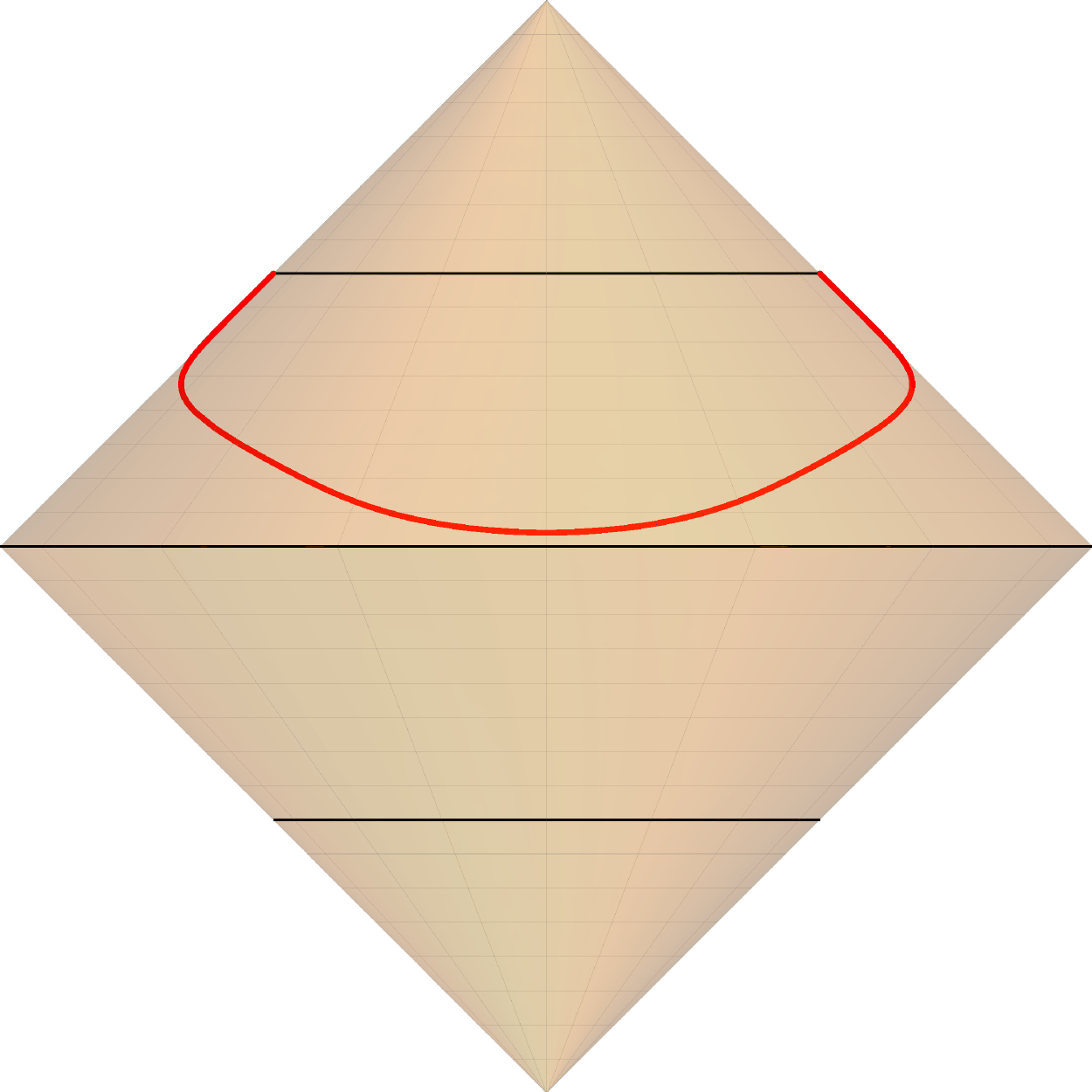}
	}
        \caption{Side view}
        \label{fig:B2X}
    \end{subfigure}
        \savebox{\largestimage}{\includegraphics[width=0.3\textwidth]{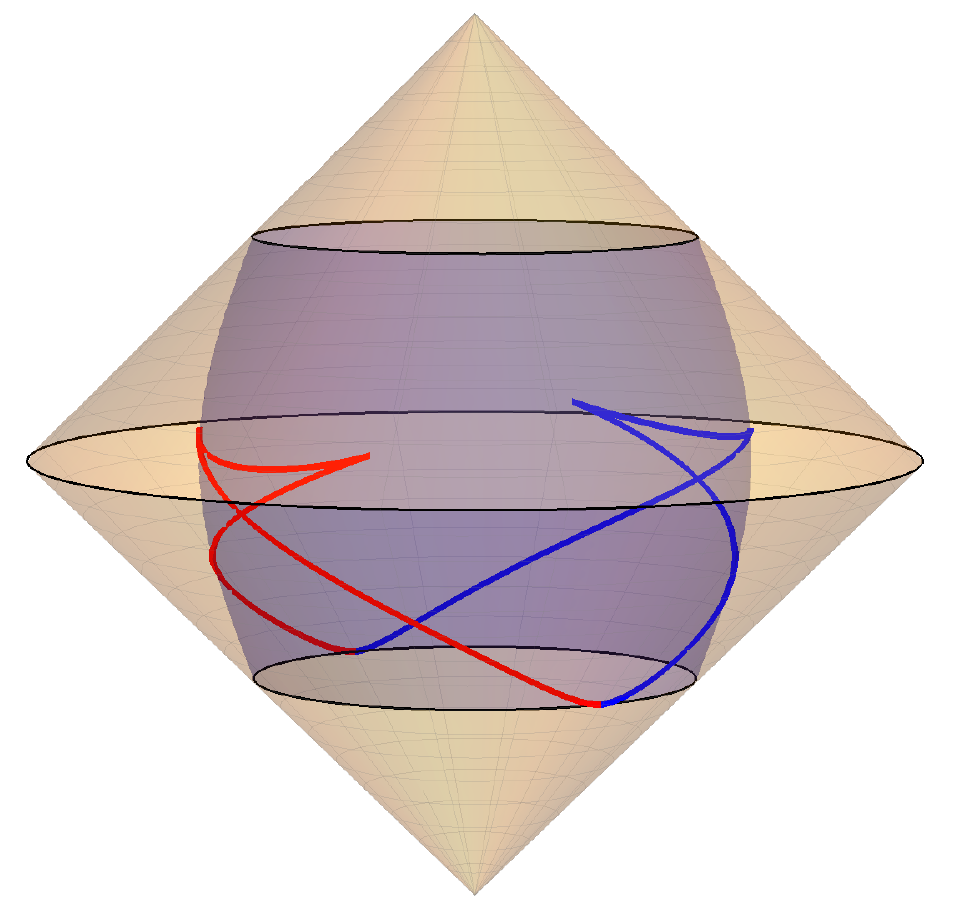}}
    \begin{subfigure}[b]{0.3\textwidth}
        \centering
        \usebox{\largestimage}
        \caption{Newman-Unti surface}
        \label{fig:NU2P}
    \end{subfigure}
    \begin{subfigure}[b]{0.3\textwidth}
        \centering
        \raisebox{\dimexpr0.5\ht\largestimage-0.5\height}{
            \includegraphics[width=0.8\textwidth]{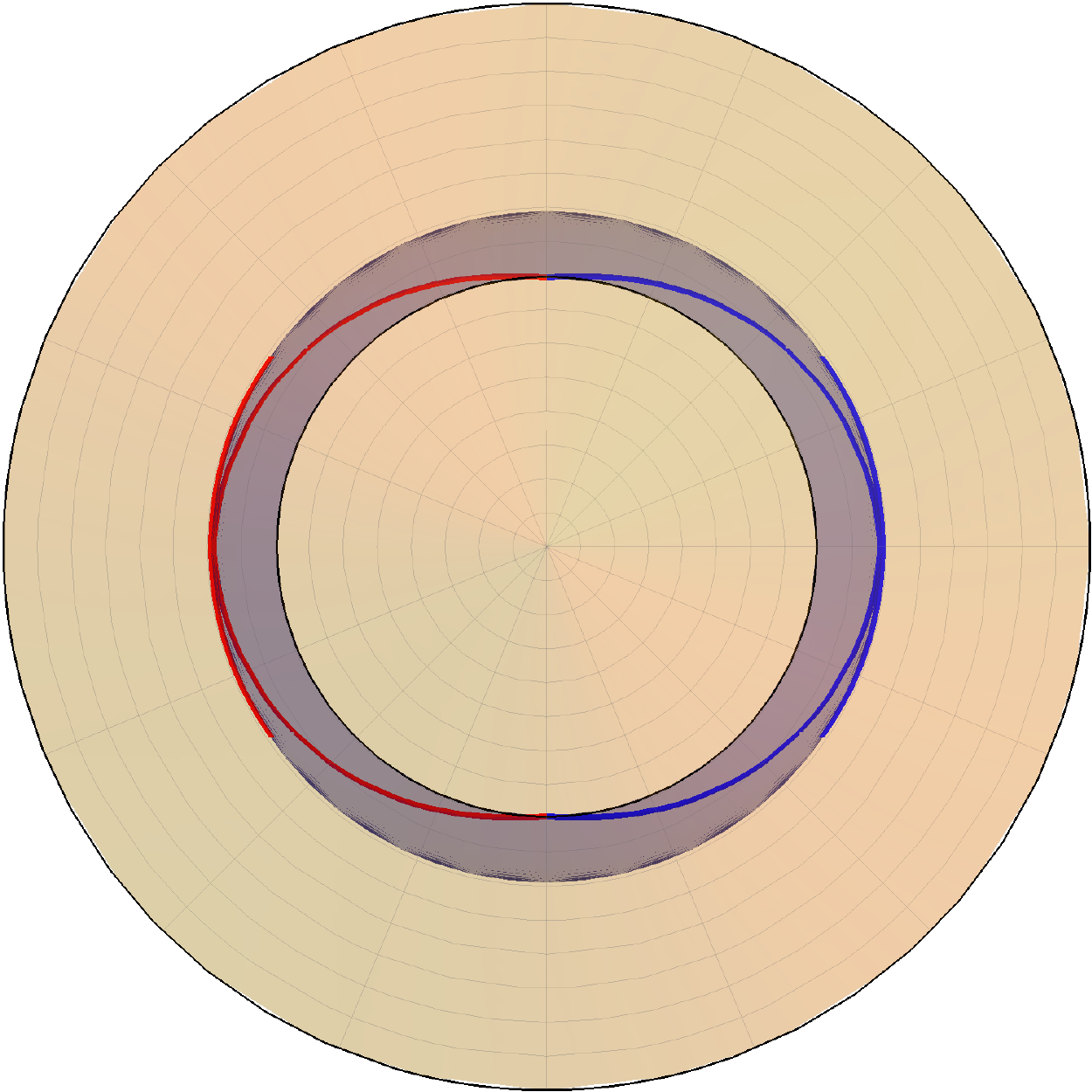}
        }
        \caption{Top view}
        \label{fig:NU2T}
    \end{subfigure}
%    \\
    \hspace{-0.2cm}
    \begin{subfigure}[b]{0.3\textwidth}
        \centering
        \raisebox{\dimexpr0.5\ht\largestimage-0.5\height}{
        		\includegraphics[width=0.9\textwidth]{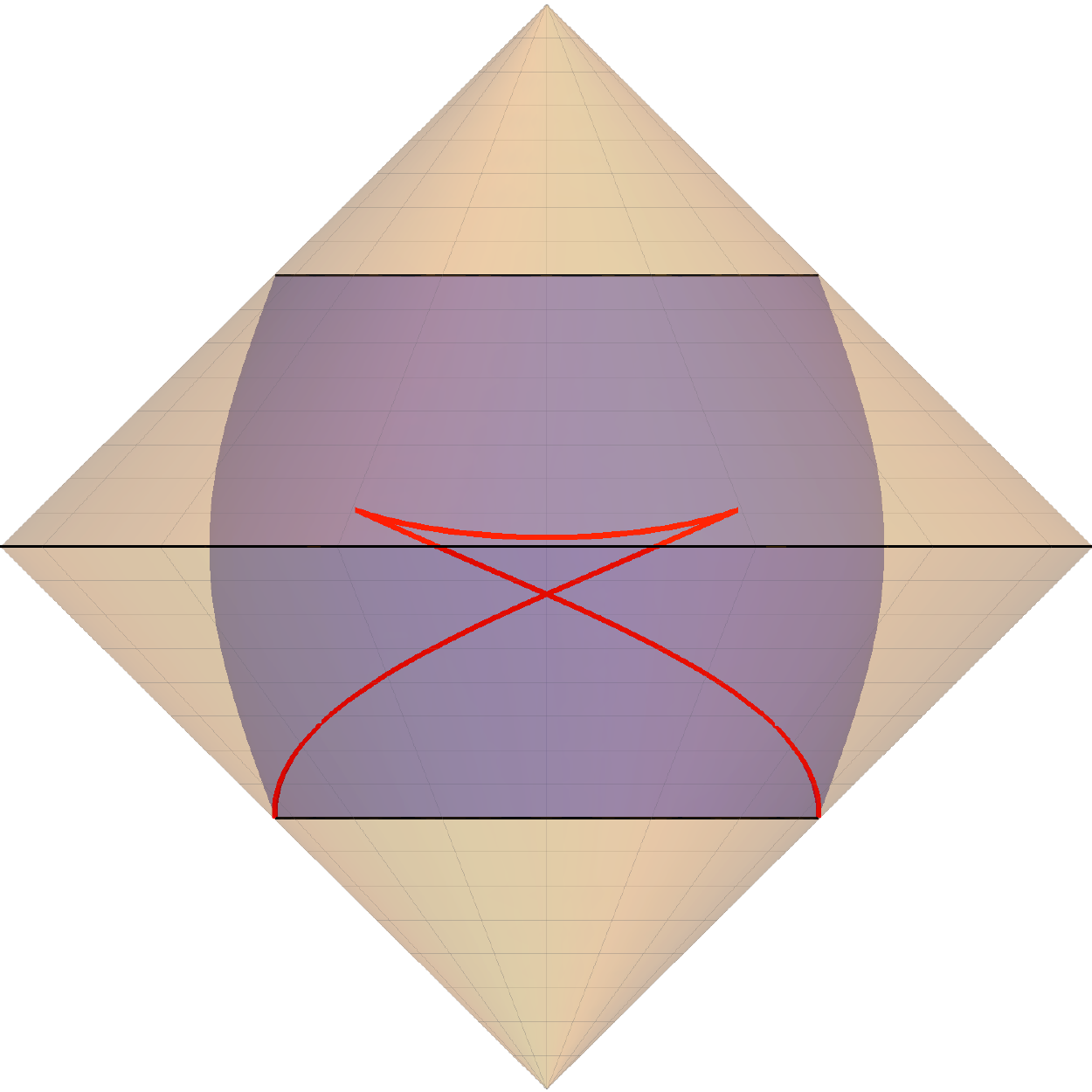}
	}
        \caption{Side view}
        \label{fig:NU2X}
    \end{subfigure}
    \caption{%$\alpha = 1.4$, $r=1.6$, $u=-1$ 
     Bondi and Newman-Unti surfaces with $\alpha>1$, $u<0$. 
     }
    \label{fig:a+u-}
\end{figure}

\begin{figure}[t]
    \centering
    \savebox{\largestimage}{\includegraphics[width=0.3\textwidth]{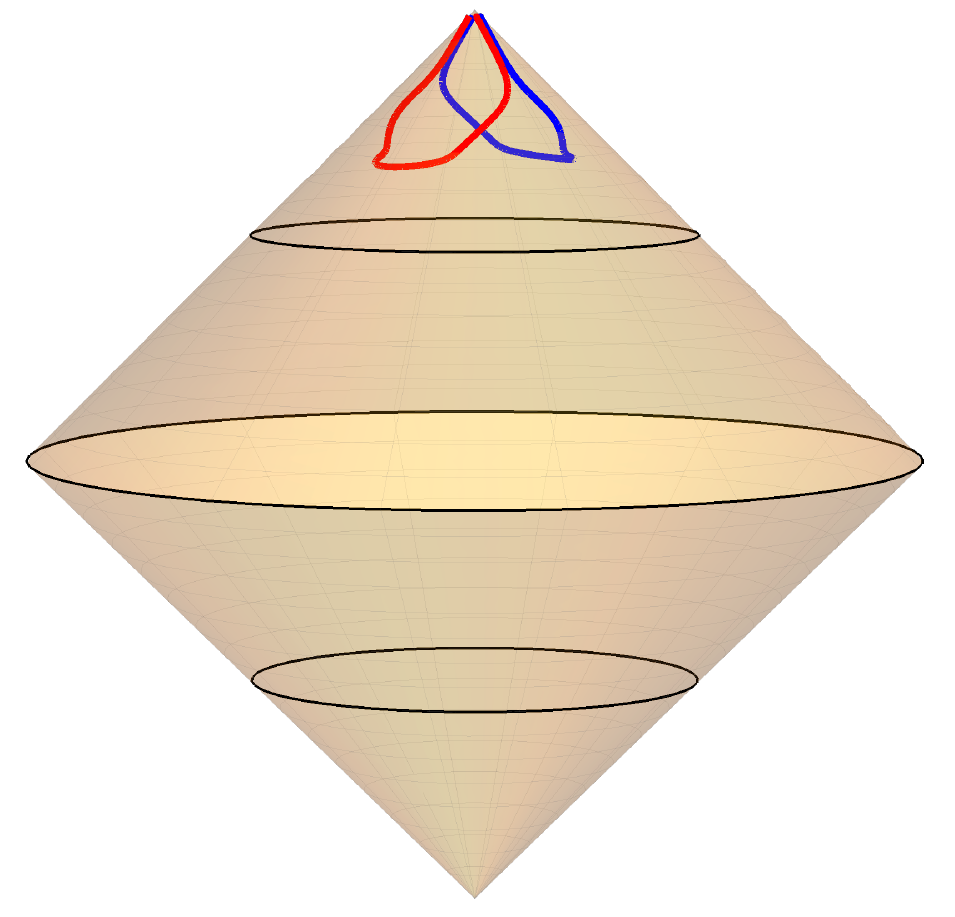}}
    \begin{subfigure}[b]{0.3\textwidth}
        \centering
        \usebox{\largestimage}
        \caption{Bondi surface}
        \label{fig:B3P}
    \end{subfigure}
    \begin{subfigure}[b]{0.3\textwidth}
        \centering
        \raisebox{\dimexpr0.5\ht\largestimage-0.5\height}{
            \includegraphics[width=0.8\textwidth]{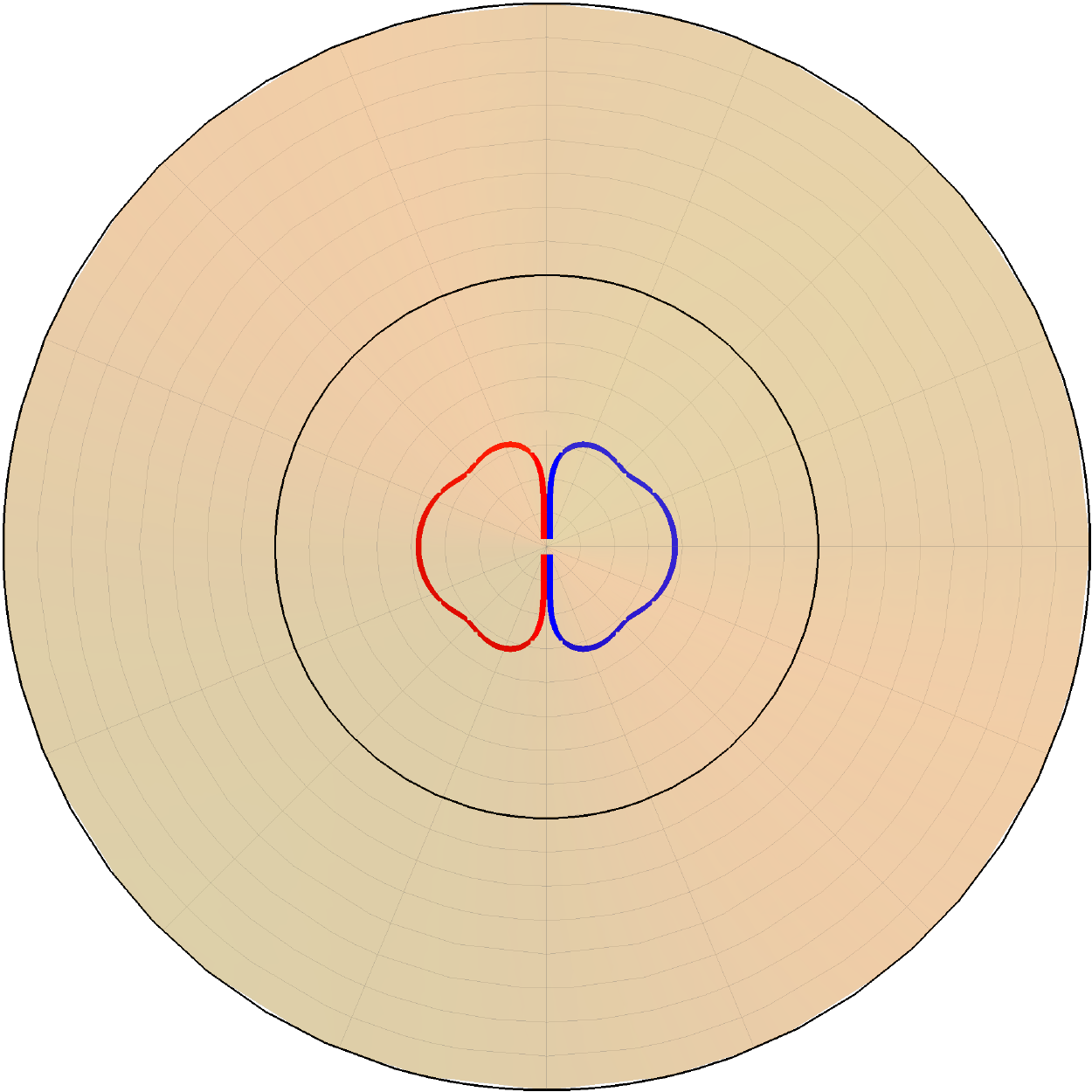}
        }
        \caption{Top view}
        \label{fig:B3T}
    \end{subfigure}
    \hspace{-0.2cm}
    \begin{subfigure}[b]{0.3\textwidth}
        \centering
        \raisebox{\dimexpr0.5\ht\largestimage-0.5\height}{
        		\includegraphics[width=0.9\textwidth]{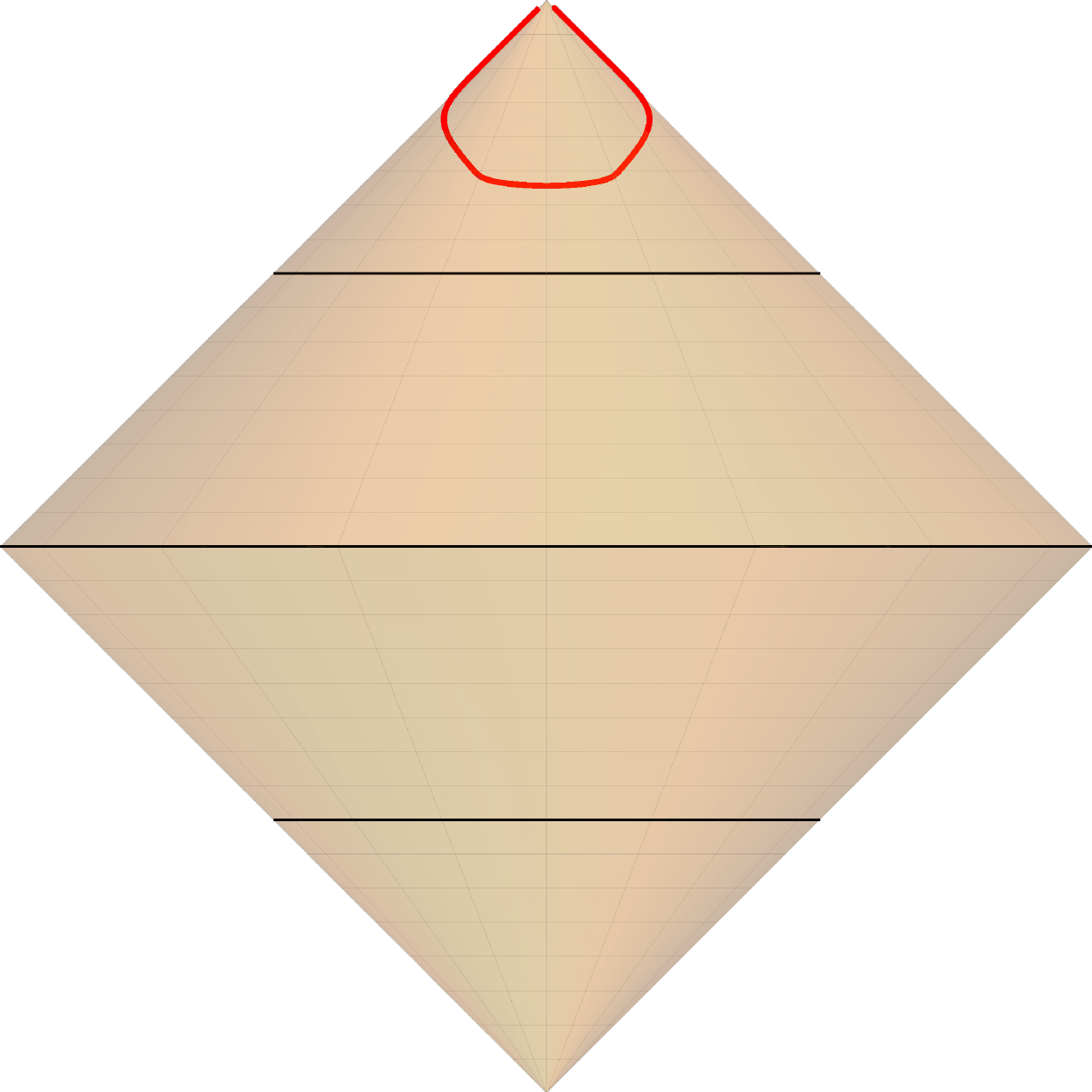}
	}
        \caption{Side view}
        \label{fig:B3X}
    \end{subfigure}
        \savebox{\largestimage}{\includegraphics[width=0.3\textwidth]{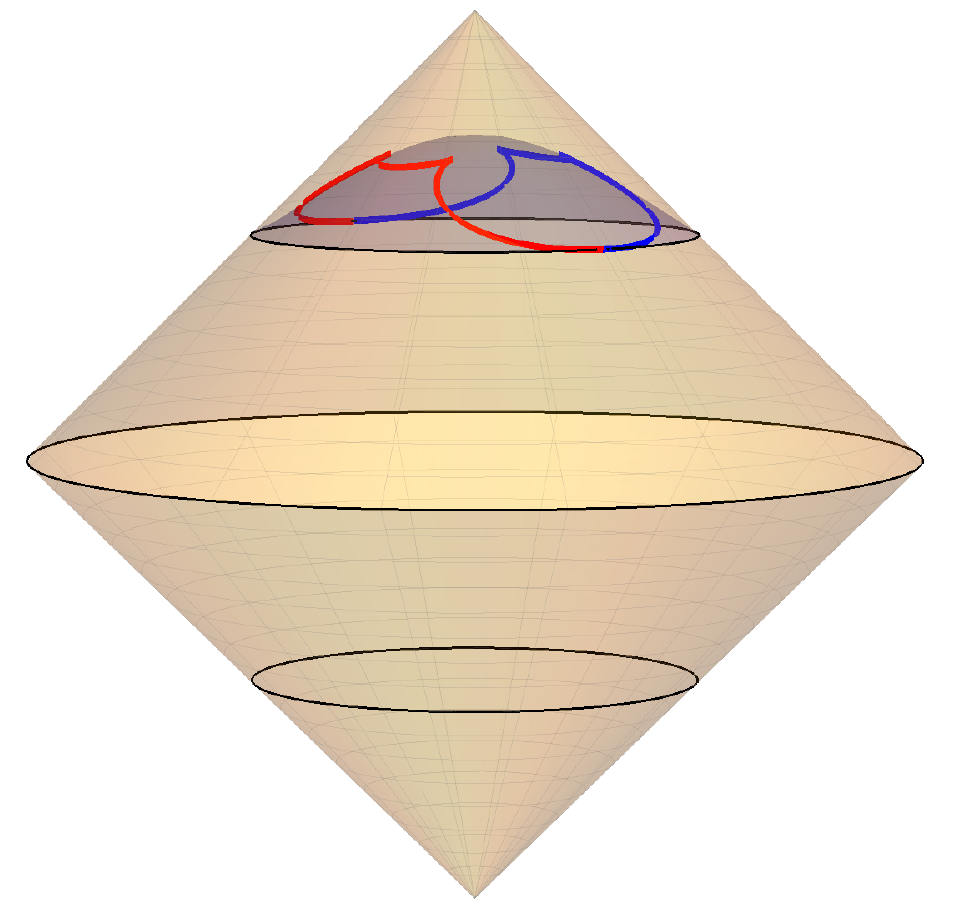}}
    \begin{subfigure}[b]{0.3\textwidth}
        \centering
        \usebox{\largestimage}
        \caption{Newman-Unti surface}
        \label{fig:NU3P}
    \end{subfigure}
    \begin{subfigure}[b]{0.3\textwidth}
        \centering
        \raisebox{\dimexpr0.5\ht\largestimage-0.5\height}{
            \includegraphics[width=0.8\textwidth]{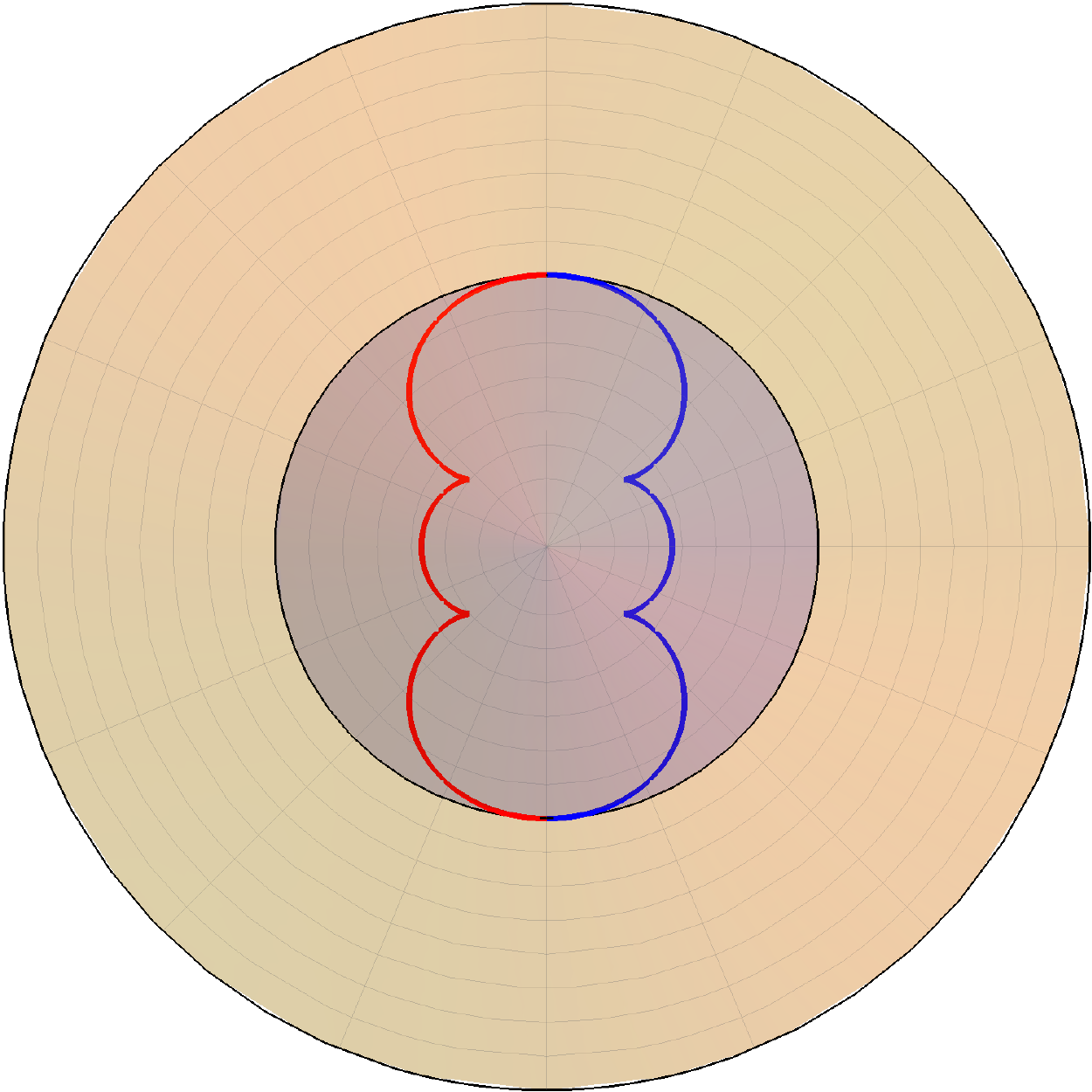}
        }
        \caption{Top view}
        \label{fig:NU3T}
    \end{subfigure}
%    \\
    \hspace{-0.2cm}
    \begin{subfigure}[b]{0.3\textwidth}
        \centering
        \includegraphics[width=0.9\textwidth]{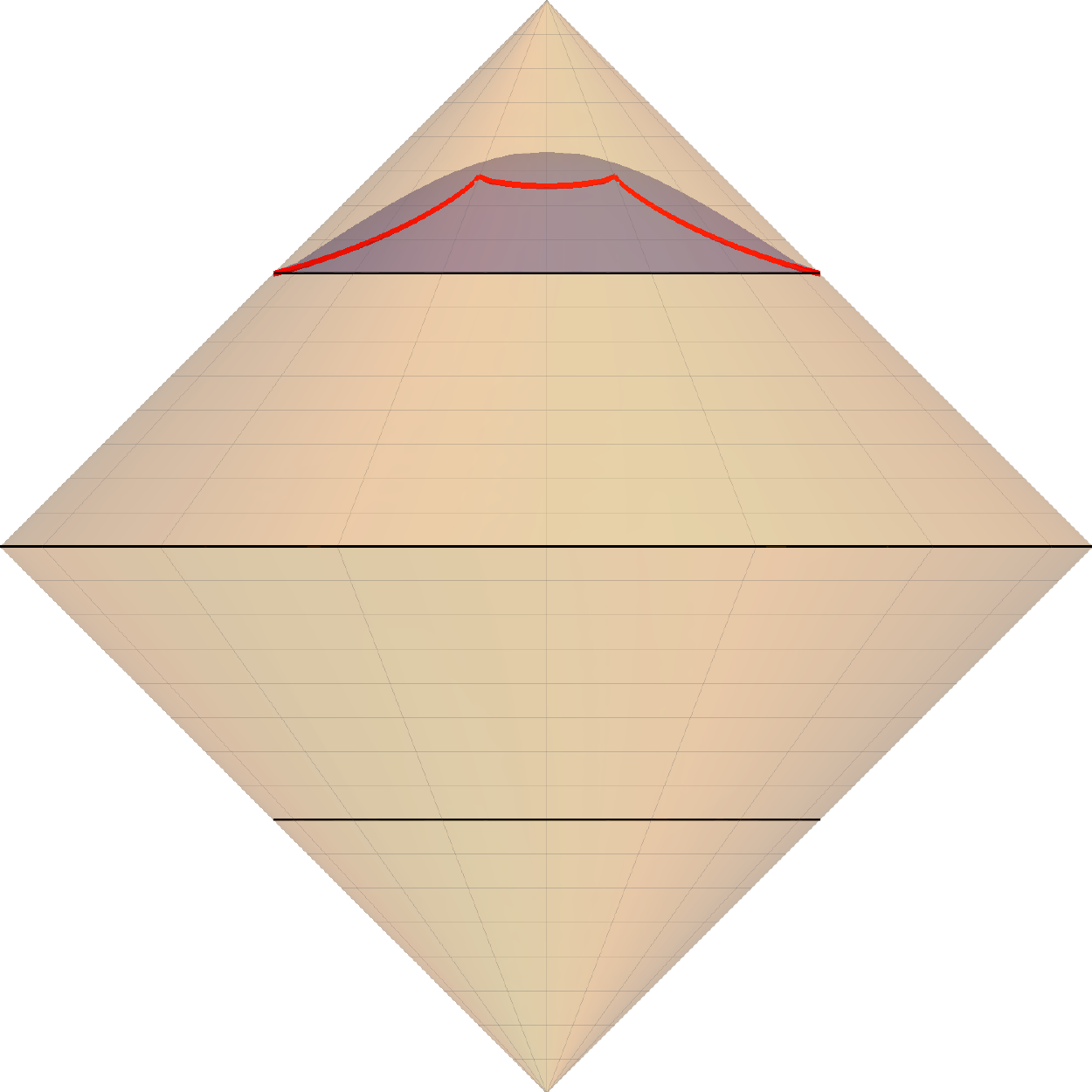}
        \caption{Side view}
        \label{fig:NU3X}
    \end{subfigure}

    \caption{%$\alpha = 0.8$, $r=2.315$, $u=1$  
     Bondi and Newman-Unti surfaces with $\alpha<1$, $u>0$. }
    \label{fig:a-u+}
\end{figure}

\begin{figure}[t]
    \centering
    \savebox{\largestimage}{\includegraphics[width=0.3\textwidth]{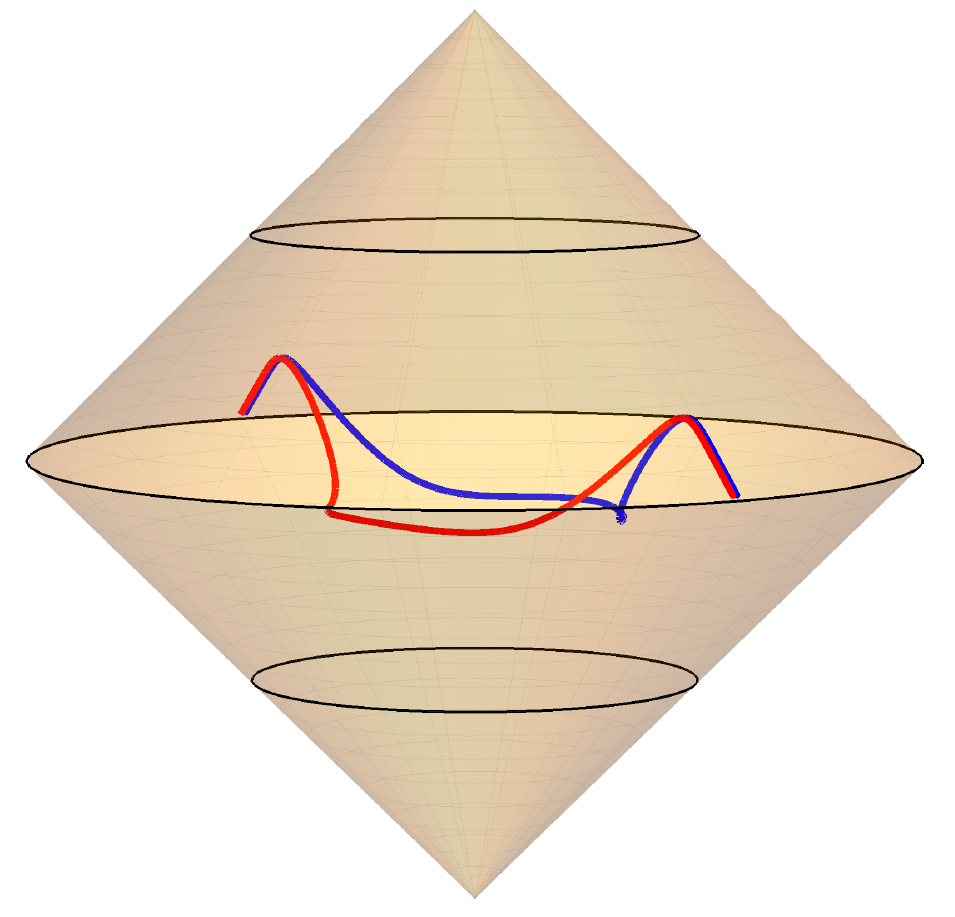}}
    \begin{subfigure}[b]{0.3\textwidth}
        \centering
        \usebox{\largestimage}
        \caption{Bondi surface}
        \label{fig:B4P}
    \end{subfigure}
    \begin{subfigure}[b]{0.3\textwidth}
        \centering
        \raisebox{\dimexpr0.5\ht\largestimage-0.5\height}{
            \includegraphics[width=0.8\textwidth]{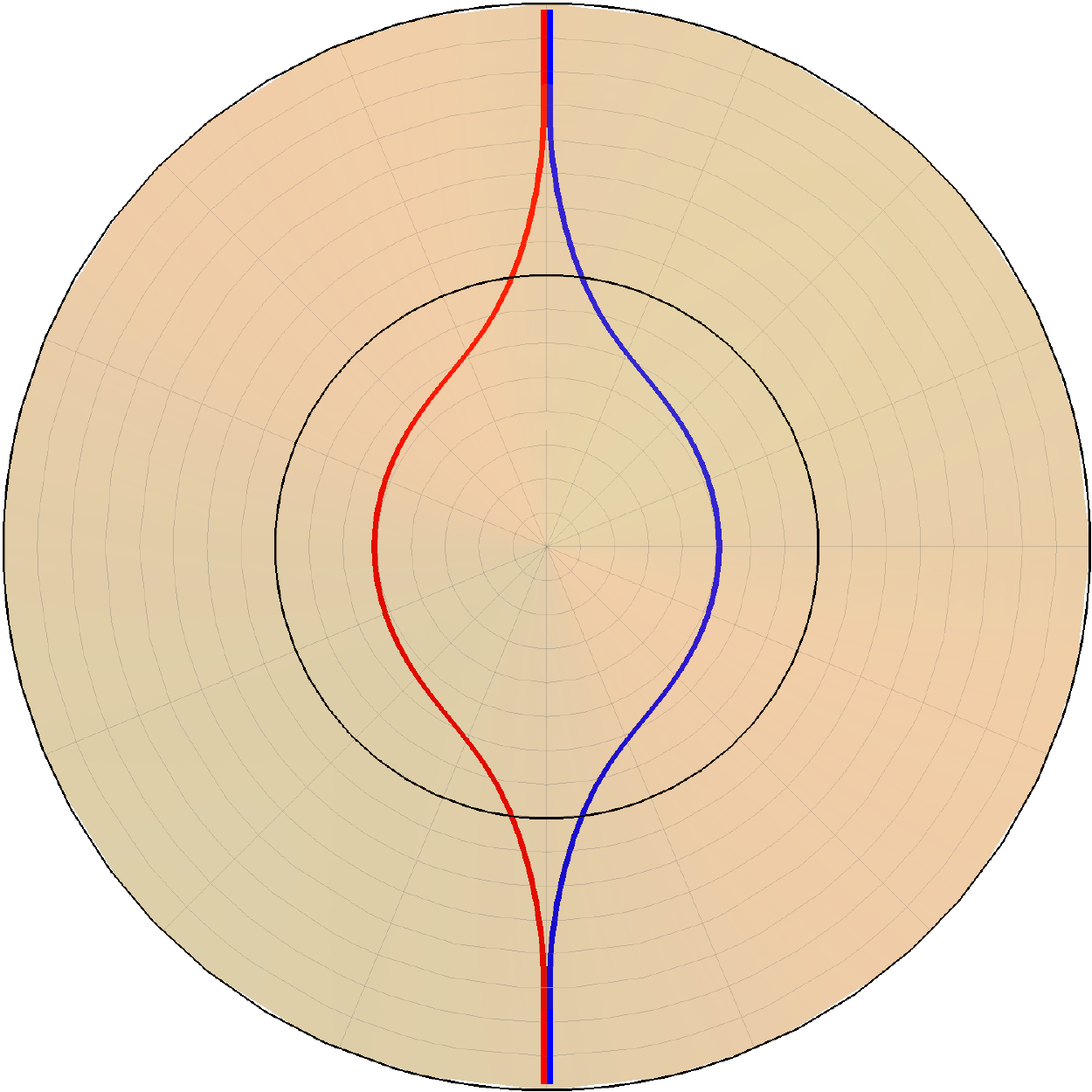}
        }
        \caption{Top view}
        \label{fig:B4T}
    \end{subfigure}
    \hspace{-0.2cm}
    \begin{subfigure}[b]{0.3\textwidth}
        \centering
        \raisebox{\dimexpr0.5\ht\largestimage-0.5\height}{
        		\includegraphics[width=0.9\textwidth]{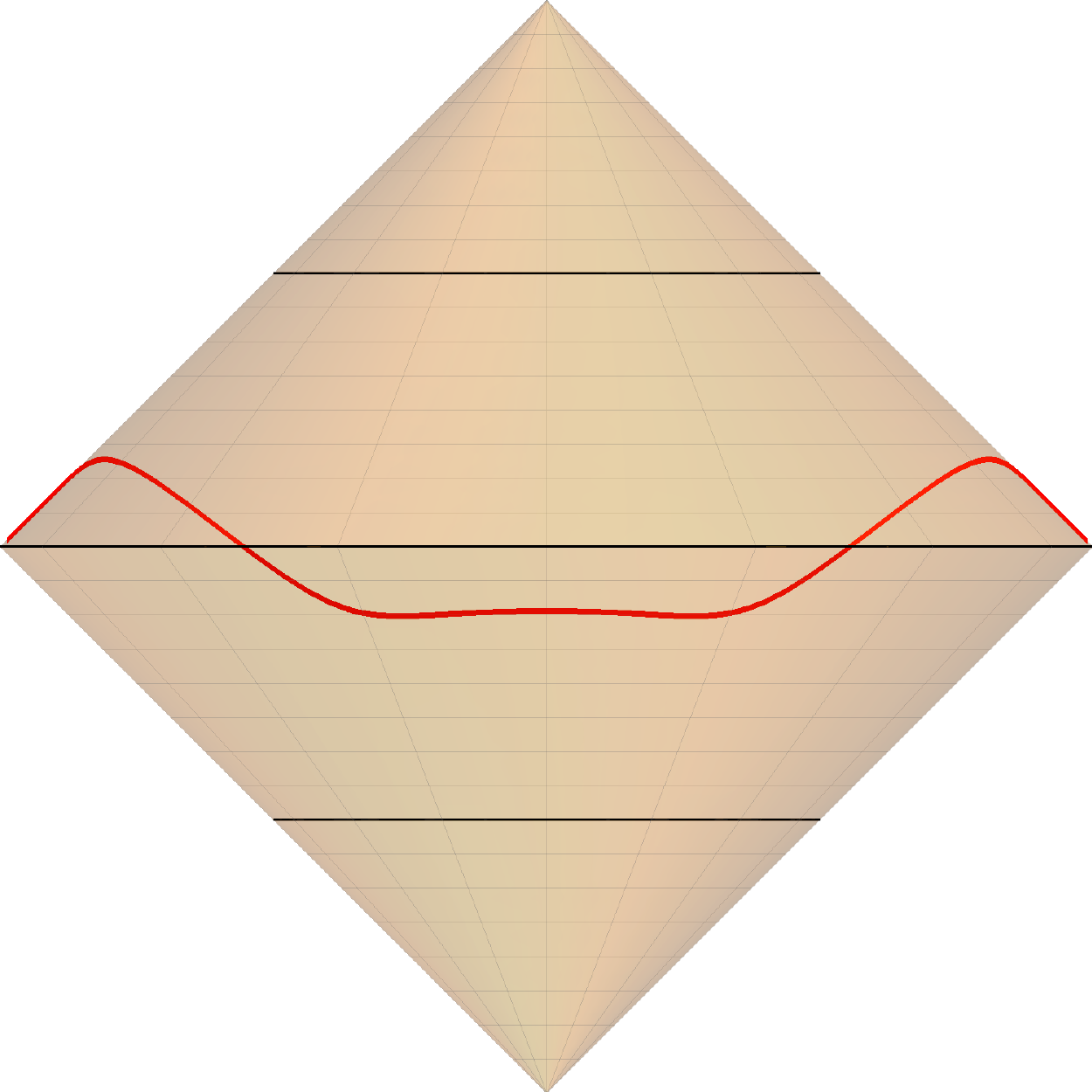}
	}
        \caption{Side view}
        \label{fig:B4X}
    \end{subfigure}
        \savebox{\largestimage}{\includegraphics[width=0.3\textwidth]{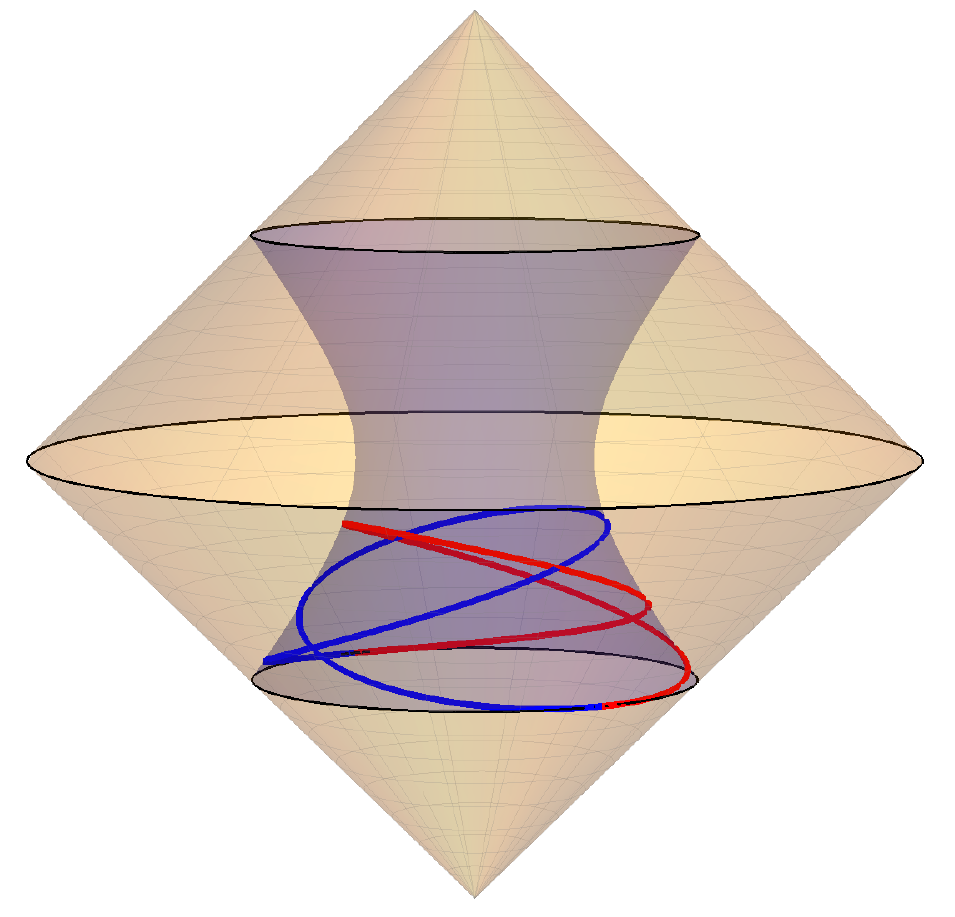}}
    \begin{subfigure}[b]{0.3\textwidth}
        \centering
        \usebox{\largestimage}
        \caption{Newman-Unti surface}
        \label{fig:NU4P}
    \end{subfigure}
    \begin{subfigure}[b]{0.3\textwidth}
        \centering
        \raisebox{\dimexpr0.5\ht\largestimage-0.5\height}{
            \includegraphics[width=0.8\textwidth]{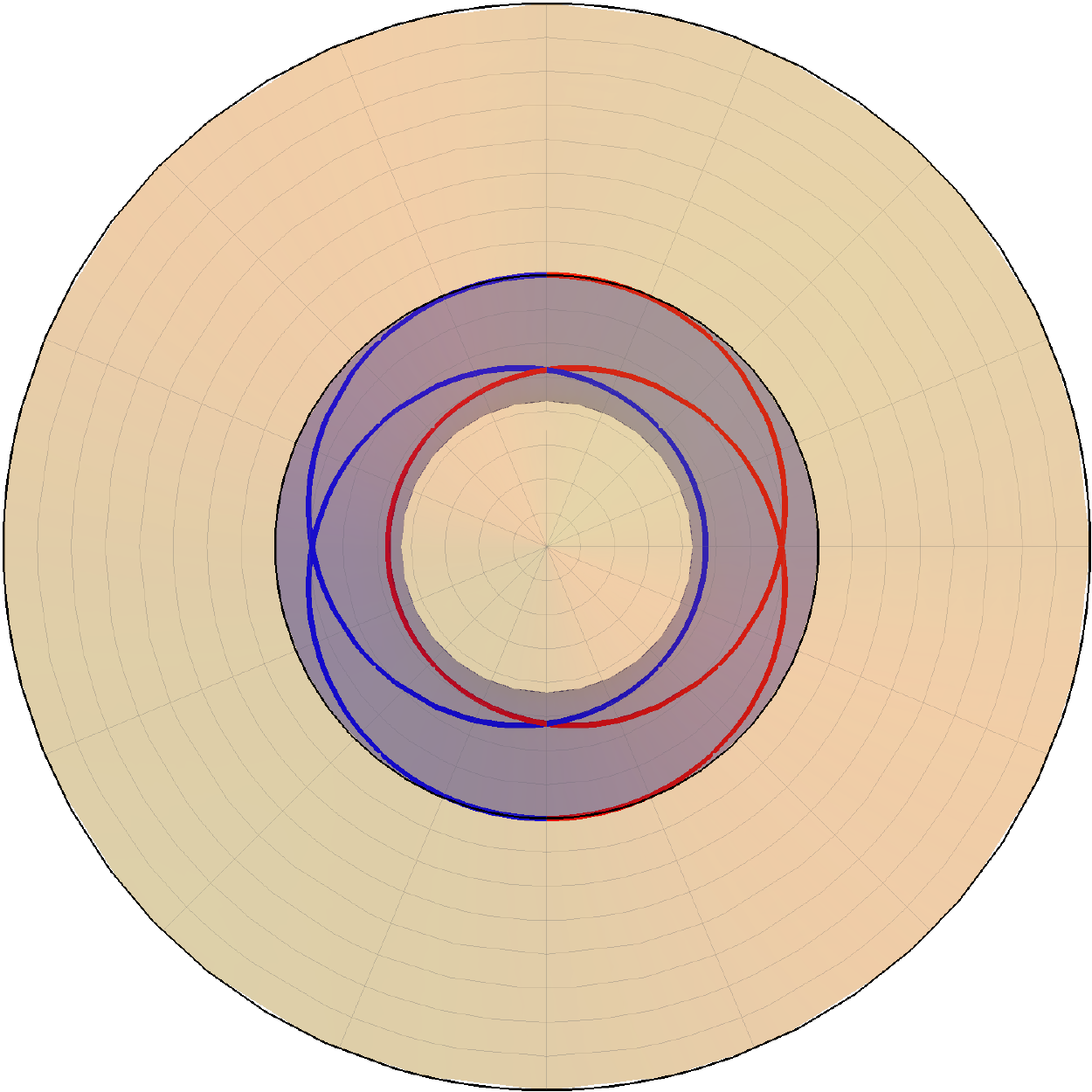}
        }
        \caption{Top view}
        \label{fig:NU4T}
    \end{subfigure}
    \hspace{-0.2cm}
    \begin{subfigure}[b]{0.3\textwidth}
        \centering
        \raisebox{\dimexpr0.5\ht\largestimage-0.5\height}{
        		\includegraphics[width=0.9\textwidth]{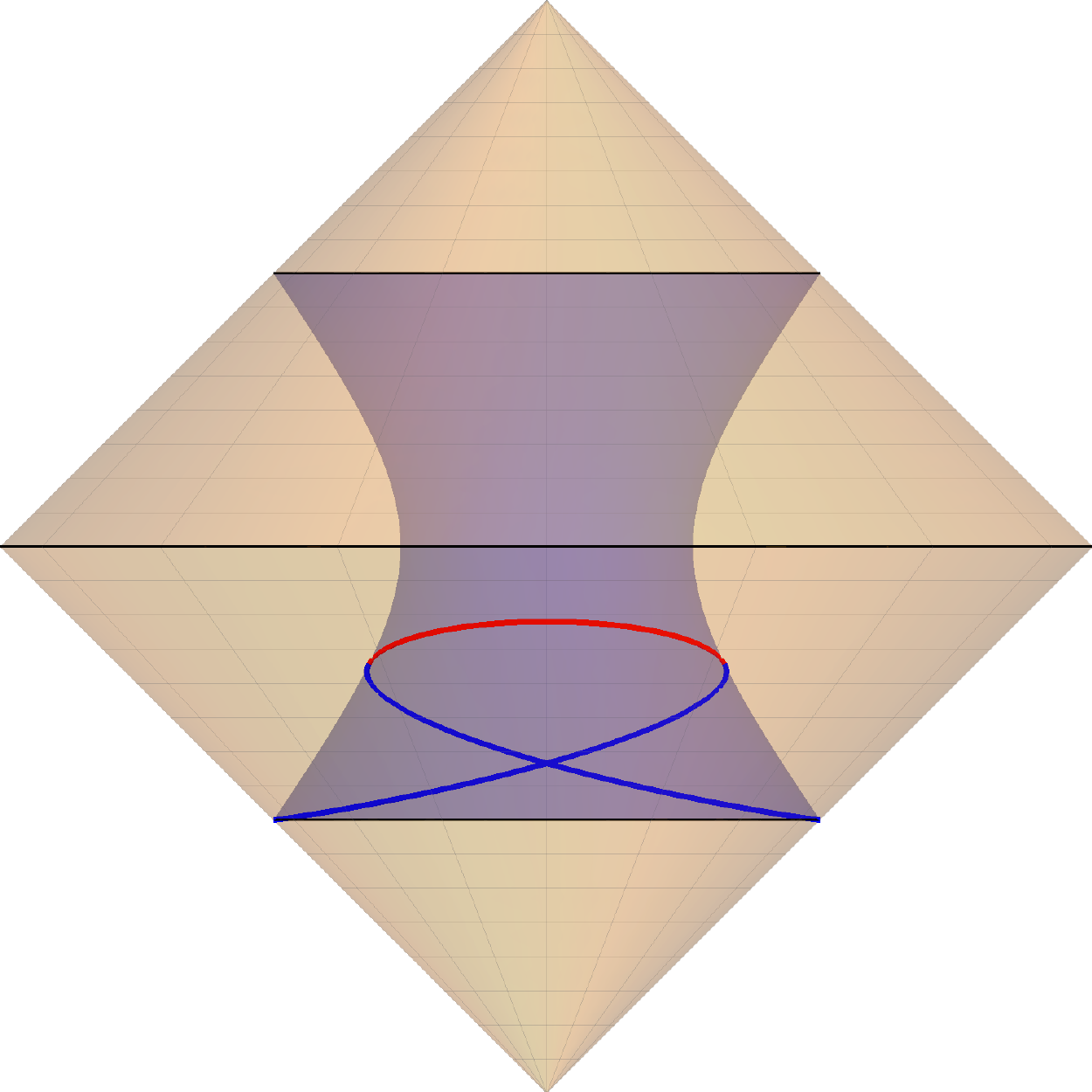}
	}
        \caption{Side view}
        \label{fig:NU4X}
    \end{subfigure}
    \caption{%$\alpha = 0.8$, $r=0.6$, $u=-1$  
     Bondi and Newman-Unti surfaces with $\alpha<1$, $u<0$.}
    \label{fig:a-u-}
\end{figure}

\begin{figure}[t]
    \centering
    \savebox{\largestimage}{\includegraphics[width=0.3\textwidth]{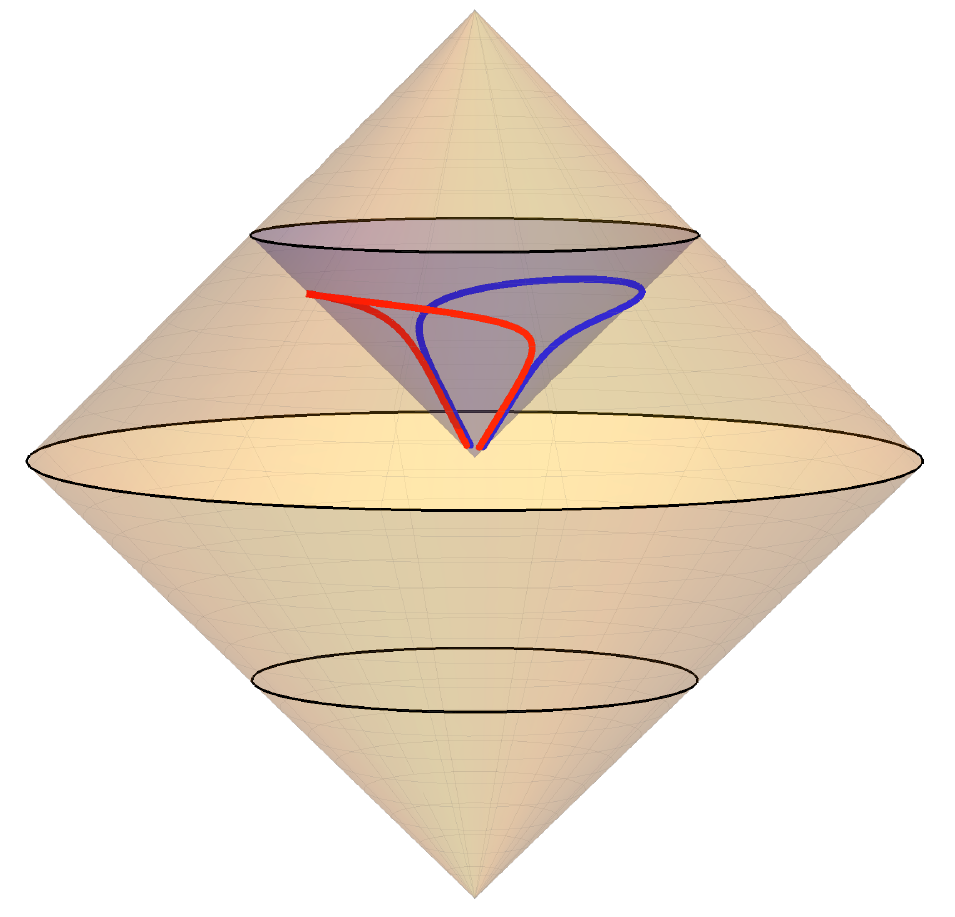}}
    \begin{subfigure}[b]{0.3\textwidth}
        \centering
        \usebox{\largestimage}
        \caption{Bondi/NU surface}
        \label{fig:NU5P}
    \end{subfigure}
    \begin{subfigure}[b]{0.3\textwidth}
        \centering
        \raisebox{\dimexpr0.5\ht\largestimage-0.5\height}{
            \includegraphics[width=0.8\textwidth]{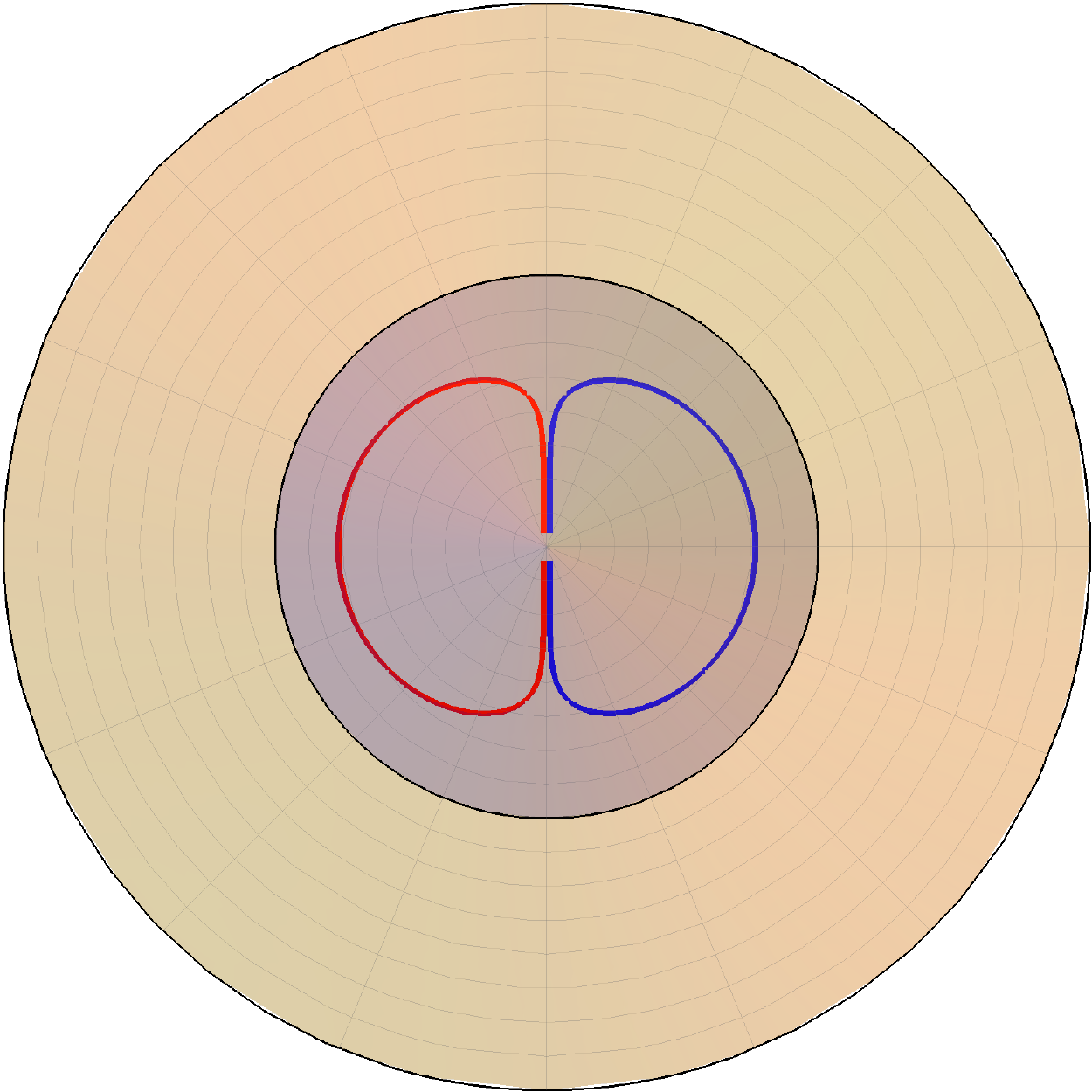}
        }
        \caption{Top view}
        \label{fig:NU5T}
    \end{subfigure}
    \hspace{-0.2cm}
    \begin{subfigure}[b]{0.3\textwidth}
        \centering
        \raisebox{\dimexpr0.5\ht\largestimage-0.5\height}{
        		\includegraphics[width=0.9\textwidth]{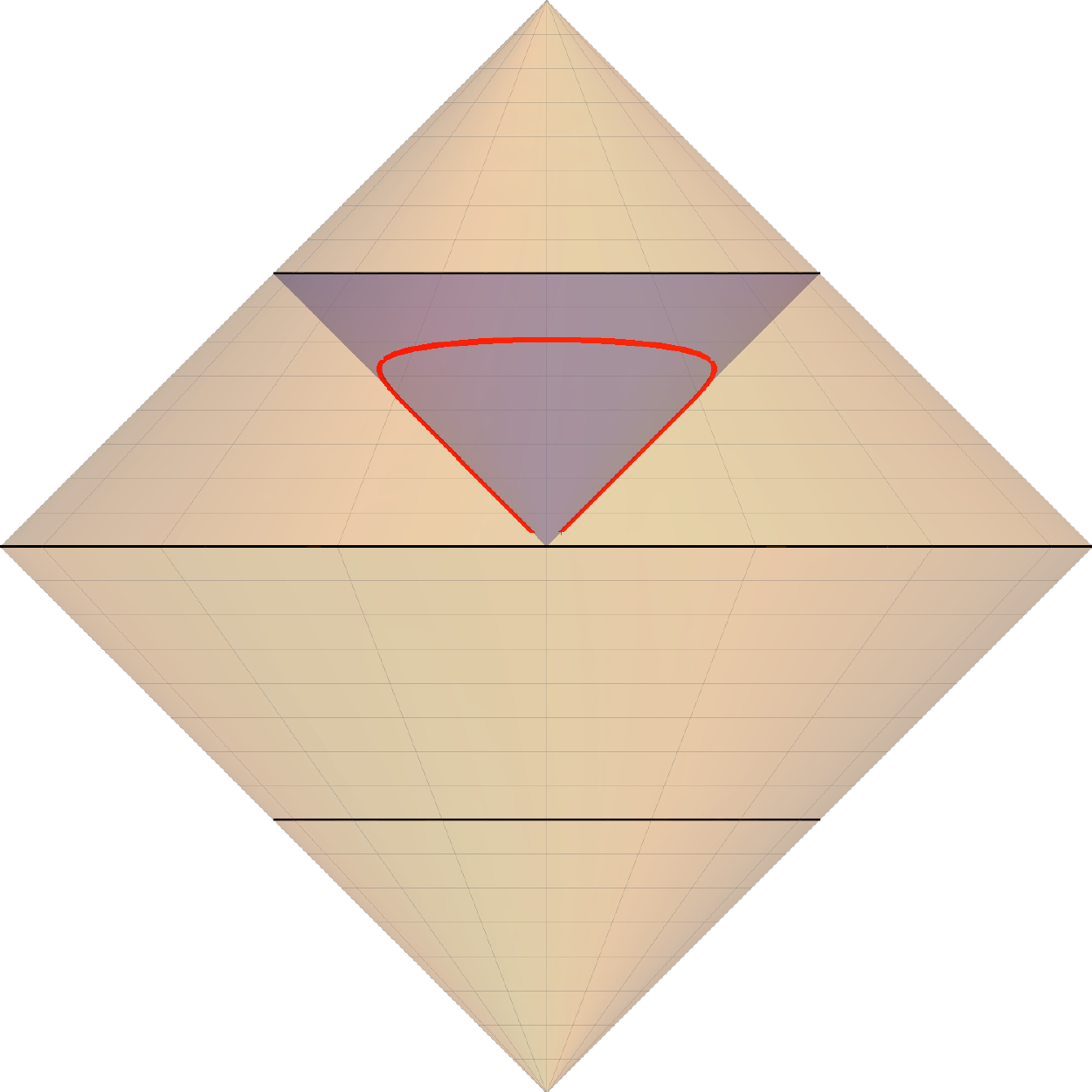}
	}
        \caption{Side view}
        \label{fig:NU5X}
    \end{subfigure}
    \caption{%$\alpha = 0.8$, $r=1$, $u=0$  
    Surface with $\alpha<1$, $u=0$. The Bondi and Newman-Unti surfaces are the 
    same in this case, since $\rho=r$ when $u=0$. }
    \label{fig:a-u0}
\end{figure}

%%%%%%

Equation (\ref{eqn:finitesuperrot}) can be used to compute a number of quantities 
which relate to how the cosmic footballs and cosmic horns embed into Minkowski space.  
One such quantity is the radial distance of the image point from the origin,
\begin{align} 
    \vec x^2 &= t^2 + 2 c t + c^2 + \vec b^2 \nonumber \\
    &= t^2 - u^2 - 2 u \rho, \label{eqn:xvec2}
\end{align}
which also determines the Lorentzian distance of the image point,
\beq\label{eqn:propdist}
-t^2+ \vec x^2 = -u^2- 2 u \rho.
\eeq
The right hand side is simply the Lorentzian distance to the origin of the initial point $(u,\rho,z,\bar z)$ for the Newman-Unti
superrotation.  This implies  an important property of the Newman-Unti superrotation, namely that it preserves the hyperboloids 
at fixed proper distance from the origin of Minkowski space.  In fact, one could instead define the superrotation to preserve these
hyperboloids as a gauge-fixing condition for the radial coordinate, and this would produce equation (\ref{eqn:tNU}) for 
$t(u,\rho, z,\bar z)$, with the fact that the this coincides with Newman-Unti gauge a happy coincidence.  
Recently, it was argued in \cite{Ball2019} that defining superrotations to preserve hyperboloids is 
desirable in applications to flat-space holography, 
and here we see that the Newman-Unti gauge is closely related to 
this choice.  
For the Bondi superrotation, where $\rho = \sqrt{r^2 + u^2 V}$, equation (\ref{eqn:propdist}) shows that the proper distance 
to the origin is not preserved.  
Instead, the Schwarzian $\{G(z),z\}$ appearing in $V$ determines the extent to which proper distance conservation is violated in Bondi gauge.  

In section \ref{sec:bondisurf}, 
we argued that the Bondi spatial surfaces must extend out to infinity as the pole of the transformation
is approached.  This can be verified explicitly using the formulas (\ref{eqn:finitesuperrot}), which can also determine
where the spheres intersect $\scri^+$.  For this, we need the behavior of $\Omega$, $V$, and $L$ at small values of $z\bar z$
for the $G(z) = z^\alpha$ transformation,
\begin{align}
    \Omega &\rightarrow \alpha (z\bar z)^{\frac{\alpha-1}{2}}  \\
    V & \rightarrow \left(\frac{\alpha^2-1}{8 z\bar z}\right)^2 \\
    L & \rightarrow \frac{(\alpha-1)^2}{8 z\bar z},
\end{align}
which then determine the behavior of $t$ in this limit,
\begin{align}
    t &\rightarrow \frac{|u||\alpha-1|}{8\alpha (z\bar z)^{\frac{\alpha+1}{2}}}\big[\alpha+1+\sgn(u)|\alpha-1| \big]
\end{align}
The term in brackets is either $2\alpha$ or $2$, so it is always positive, and hence $t\rightarrow +\infty$ near the pole.  
The limiting value of $\rho$ is
\beq
\rho\rightarrow |u| V^{1/2} = |u| \frac{|\alpha^2-1|}{8z\bar z}.
\eeq
With these expressions, 
we can  examine the value of the retarded time $u_f = t-\sqrt{\vec x^2}$ at the pole to determine the point at which the 
surface intersects $\scri^+$.  Since $t^2$ is  much larger than $\rho$ near the pole, application
of equation (\ref{eqn:xvec2}) leads to the limiting behavior,
\begin{align}
    u_f\rightarrow \frac{u\rho}{t} \sim u(z\bar z)^{\frac{\alpha-1}{2}}.
\end{align}
When $\alpha>1$, this goes to $0$ at small $z\bar z$, and so the Bondi sphere intersects $\scri^+$ on the celestial
sphere at $u_f=0$.  On the other hand, when $\alpha<1$, this quantity diverges to either $\pm \infty$, depending on the sign
of $u$.  
For positive values of $u$, the spheres approach timelike infinity with $u_f\rightarrow+\infty$, while for negative values of $u$, they approach spatial infinity with $u_f\rightarrow -\infty$.

This behavior is confirmed in figures \ref{fig:a+u+}, \ref{fig:a+u-}, \ref{fig:a-u+}, and \ref{fig:a-u-},
which plot the embedding of Bondi surfaces 
into conformally compactified Minkowksi space\footnote{The compactified coordinates 
we use are obtained by defining 
$
%\begin{align}
U = \arctan\left(t-\sqrt{\vec x\cdot\vec x}\right) %\\
$,
$
W = \arctan\left(t+\sqrt{\vec x \cdot\vec x}\right)
%\end{align}
$,
and then setting the compactified Cartesian coordinates to 
$
T=\frac{W+U}{2} 
$,
$
\vec X= \frac{W-U}{2} \frac{\vec x}{\sqrt{\vec x \cdot \vec x}}
$.}
for various values of $\alpha$ and $u$.  In addition, these 
figures include plots of surfaces with fixed Newman-Unti radial coordinate $\rho$, at the 
same values of $u$ and $\alpha$.  Since Newman-Unti surfaces lie on a hyperboloid
at fixed distance to the origin, this hyperboloid is also plotted in blue in the figures.  Near the 
equator $\theta = \pi/2$, the Bondi and Newman-Unti surfaces are roughly spherical, and 
their embeddings resemble each other.  As the poles are approached, the embeddings
become dissimilar and deviate from the spherical shape.  The Bondi surfaces extend out to 
infinity and tend to become tangent to $\scri^+$ as they asymptote to either the $u_f=0$ 
celestial sphere or $i^+$ or $i^0$ according to the previous discussion.  By contrast, 
the Newman-Unti surfaces develop kinks at a finite value of $\theta$, which corresponds
to the point where the induced metric (\ref{eqn:NUmetric}) becomes singular.  These kinks
lie at the caustics of the radial null congruence, and the points between the kinks and the 
poles are in a region of Minkowski space not covered by the Bondi coordinate patch.  
The singularities in the Newman-Unti spheres exhibit a variety of interesting topologies, 
including a pair of cusps in figure \ref{fig:a-u+}, a swallowtail in figure \ref{fig:a+u-},  self-intersections
in figure \ref{fig:a+u+}, and loops in figure \ref{fig:a-u-}.  Figure \ref{fig:a-u0} depicts the special
case of a surface at $u=0$, for which the Bondi and Newman-Unti surfaces coincide.  Unlike
the other cases considered, this surface does not intersect infinity at the pole when $\alpha<1$, but
rather approaches the origin of Minkowski space.  

A final surprising property of the Newman-Unti surfaces is that when $u < 0$ they intersect $\scri^-$ at the poles, as is evident in figures \ref{fig:a+u-} and \ref{fig:a-u-}.
This can be understood from \eqref{eqn:tNU}: $L$ diverges at the poles, so $t \to -\infty$ as $\theta \to 0,\pi$ with $\rho$ fixed.
Since the embedding lies on a fixed hyperboloid, these points hit $\scri^-$.
This is in sharp contrast with the limit where $\theta$ is fixed and $\rho \to \infty$, which hits $\scri^+$.
This gives a rather explicit illustration of the breakdown of the large $\rho$ limit for small $\theta$.

\subsection{Extension beyond Bondi coordinate patch}
\label{sec:extension}

\begin{figure}
    \centering
    \begin{subfigure}[t]{0.49\textwidth}
        \includegraphics[width=\textwidth]{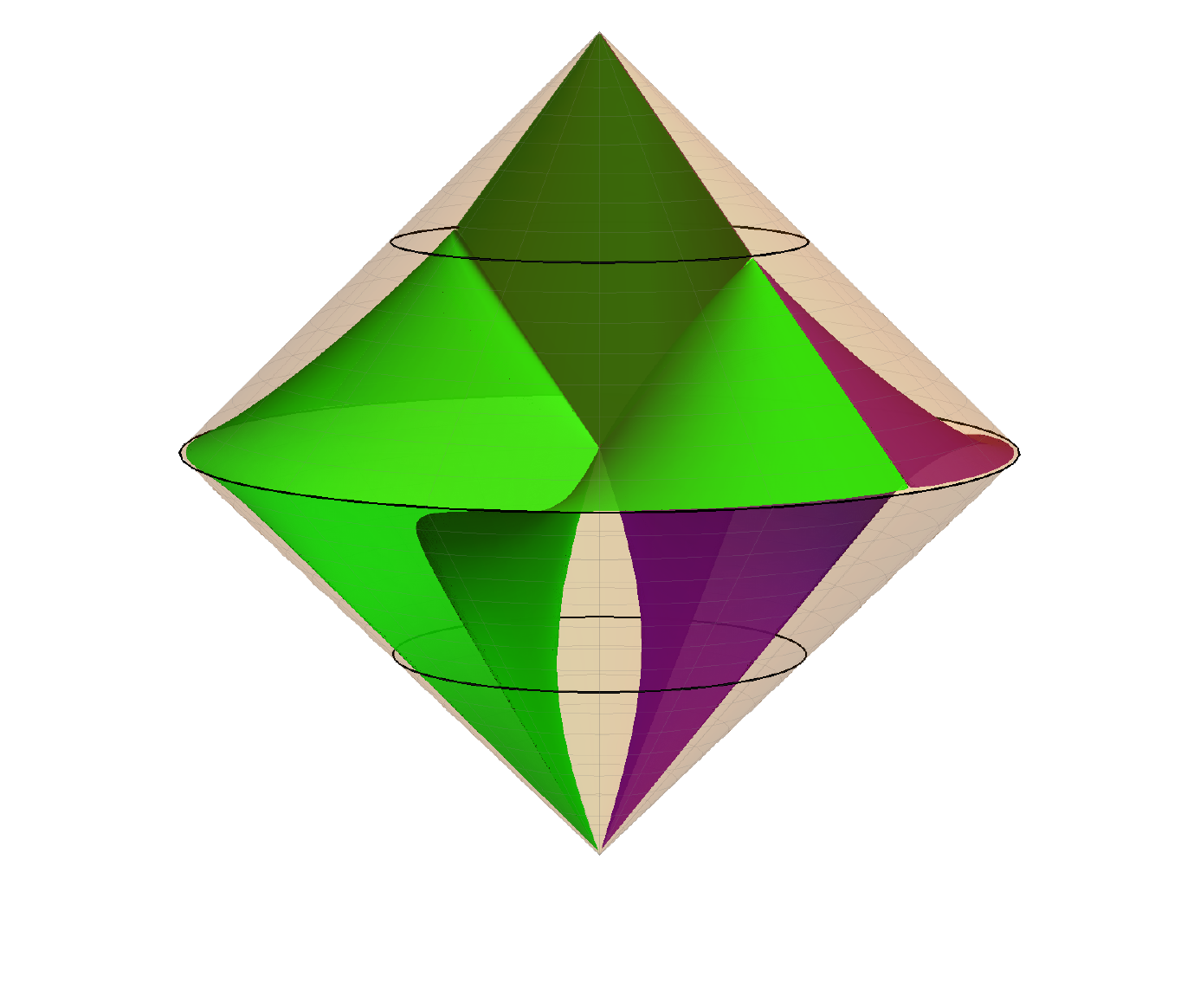}
        \caption{$\alpha>1$}
        \label{fig:a>1bh}
    \end{subfigure}
    \begin{subfigure}[t]{0.49\textwidth}
        \includegraphics[width=\textwidth]{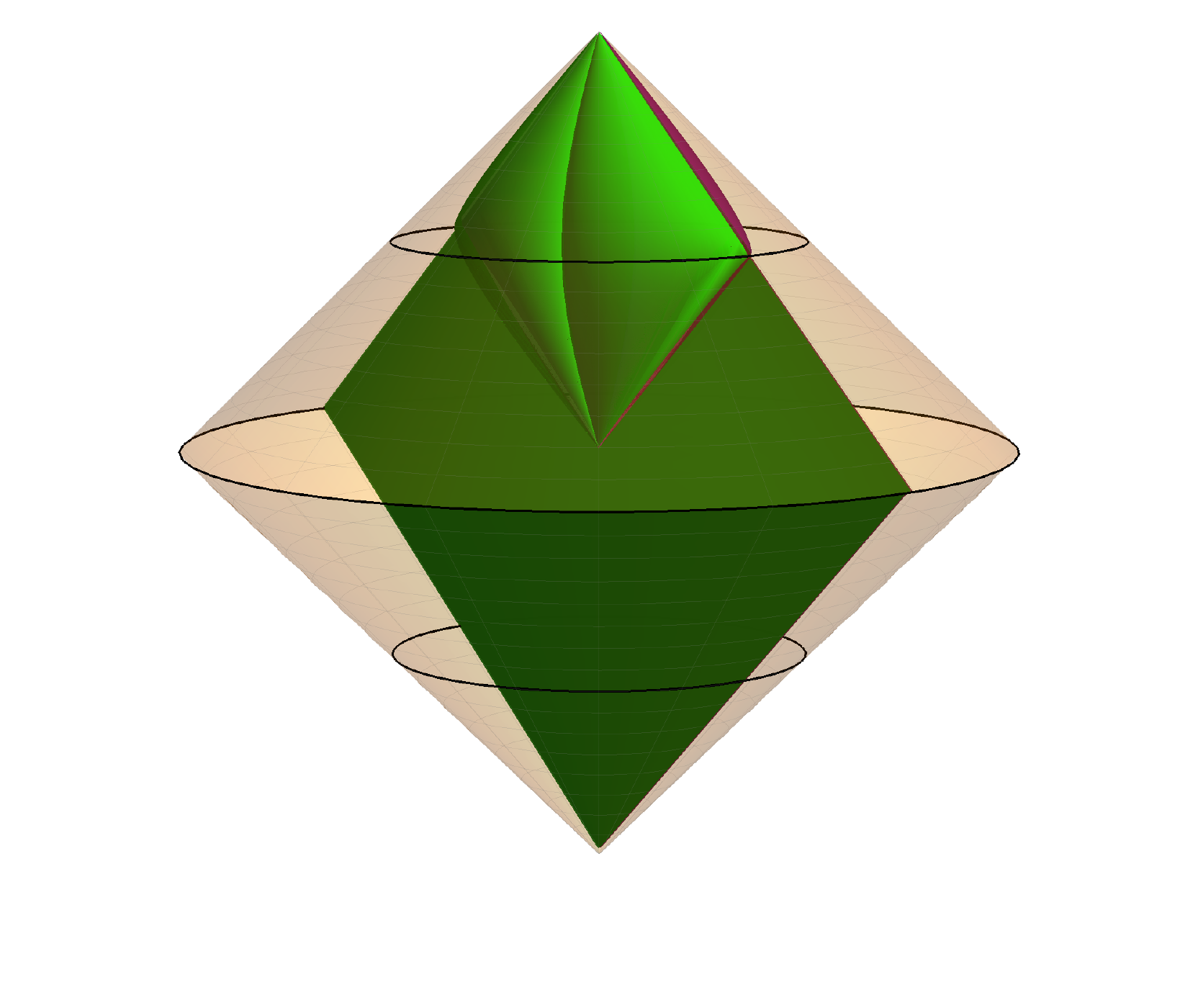}
        \caption{$\alpha<1$}
        \label{fig:a<1bh}
    \end{subfigure}
    \caption{These plots show the 
    $r=0$ hypersurfaces denoting the boundary of the Bondi coordinate patch in compactified
    Minkowski space for either sign of $\alpha-1$.  
    The $\phi$ coordinate is suppressed due to rotational symmetry, and the 
    green and purple surfaces denote antipodal points at $\phi=0$ and $\phi=\pi$.  }
    \label{fig:r=0surf}
\end{figure}

Another use for equations (\ref{eqn:finitesuperrot}) is to investigate the $r=0$ hypersurface where the Bondi coordinate
system breaks down.  The Bondi coordinates are associated with a congruence of null geodesics, which in generic situations 
undergo focusing and form caustics.  Since the Bondi radial coordinate $r$ is tied to the area element of a cross-section of
the congruence, it is designed to go to zero at caustics.  The $r=0$ surface is the boundary of 
the region covered by the Bondi coordinates, and it is of interest to know how much of Minkowski space lies beyond this 
surface.  (\ref{eqn:finitesuperrot}) can be used to plot the $r=0$ hypersurface, which is done in figure 
\ref{fig:r=0surf}
for the $z^\alpha$ transformation.  Two qualitatively different behaviors for this hypersurface are observed, depending on the 
sign of $\alpha-1$. For $\alpha>1$, the $r=0$ surface lies along the axis of the $\phi$ rotation symmetry for $u>0$, so that it 
degenerates into a 2-dimensional surface.  The degeneration is visible in figure \ref{fig:r=0surf}
by the fact that the green and purple sheets coincide in this region.  
This surface can be thought of as the worldsheet of a cosmic string running through
the center of the space, although, as discussed above, no physical string is present if the $\phi$ periodicity is chosen 
appropriately.  For $u<0$, the hypersurface is nondegenerate, and there is an open region of Minkowski space lying behind
this surface.  The behavior for the $\alpha<1$ hypersurface is just the opposite: 
the surface degenerates when $u<0$, and when 
$u>0$ there is a region beyond the $r=0$ hypersurface.  

To describe the region beyond the $r=0$ hypersurface, the Newman-Unti coordinates come in handy.  Since the Newman-Unti radial 
coordinate $\rho$ is an affine parameter, it continues to decrease monotonically past where the Bondi coordinates break down
at $\rho = \sqrt{u^2 V}$.  In particular, equation (\ref{eqn:tNU}) holds for all values of $\rho$, including $\rho<\sqrt{u^2V}$, 
which then determines the superrotation transformation in this region as well.  The spatial metric in Newman-Unti gauge for 
a general superrotation is always of the form (see equations (\ref{eqn:gNUcomp}))
\beq \label{eqn:gABNU}
g_{z\bar z} = \gamma(z,\bar z)(\rho^2 + u^2 V), \quad g_{zz} = -\rho u \{G,z\}, \quad g_{\bar z \bar z} = - \rho u \{\bar G, \bar z\}
\eeq
at any value of $\rho$.  To convert back to a Bondi coordinate system in the region $\rho<\sqrt{u^2 V}$, one simply computes the 
spatial metric determinant and sets it equal to $-r^4 \gamma(z,\bar z)^2$, which determines the relation between $\rho$ and $r$.  
This leads to the equation
\beq
r^4 = (\rho^2 - u^2 V)^2,
\eeq
whose solutions are 
\beq
r = \pm\sqrt{\pm(\rho^2-u^2 V)}, 
\eeq
and the signs are chosen to keep $r$ real, and, optionally, to make $r$ a continuous function of $\rho$.  The relations this 
produces are 
\beq
r = 
\begin{cases} 
\sqrt{\rho^2-u^2 V} & \rho^2>u^2 V  \\
-\sqrt{u^2 V - \rho^2} & \rho^2< u^2 V
\end{cases}
\eeq
\beq \label{eqn:rhor}
\rho = 
\begin{cases}
\sqrt{r^2 + u^2 V} & r>0, \rho>0 \\
\sqrt{u^2 V -r^2} & r<0, \rho>0 \\
-\sqrt{u^2 V - r^2} & r < 0, \rho<0 \\
-\sqrt{r^2+u^2V} & r>0, \rho<0
\end{cases}.
\eeq
Substituting (\ref{eqn:rhor}) into (\ref{eqn:tNU}) determines the transformation in regions beyond where the original $r>0$ Bondi 
coordinates covered.  However, as a cautionary note, the coordinate system in this region can be irregular since constant $(u,r)$ 
surfaces in this region may contain self-intersections that developed at the $r=0$ caustic.  Furthermore, the coordinates 
in this region may cover portions of Minkowski space that were already covered in the orginal Bondi patch.  This is the case 
for the $z^\alpha$ transformation in regions where the $r=0$ 
hypersurface degenerates to a two-dimensional surface, since points on
the
other side of the surface were already accounted for.

\section{Discussion}
\label{sec:discussion}

Superrotations are transformations of four-dimensional asymptotically flat spaces that arise from local conformal isometries of the celestial sphere.  
All but a finite-dimensional subgroup contain singular points and branches, and these produce unexpected features when the flow is integrated to a finite transformation of the four-dimensional geometry.  
Near poles in the transformation, the large $r$ 
asymptotic expansion breaks down, and the exact expression for the metric components is needed
to obtain reliable geometric information in this region.  In the case of Minkowski space, 
the exact result was derived by Comp\`ere and Long \cite{Compere2016b, Compere:2016jwb}, 
and the
present work gave an alternative derivation of the transformation, summarized in
equations (\ref{eqn:nullgeo}) and (\ref{eqn:finitesuperrot}), as well as 
a generalization to Newman-Unti gauge. 

The class of superrotations generated by the functions $G(z) = z^\alpha$ were examined 
in detail throughout this work.  This was motivated by the claim of Strominger and Zhiboedov
that such transformations send Minkowski space to a spacetime containing a cosmic string
\cite{Strominger:2016wns}.  We confirmed the presence of such a defect in section \ref{sec:defect} 
using the Comp\`ere-Long metric (\ref{eqn:zalphamet})
to compute the holonomy around a circle at fixed $(u,r,\theta)$.  
The defect is present if the $\phi$ coordinate is required to have 
period $2\pi$; however, when the period is set to  $\frac{2\pi}{\alpha}$, the defect is absent.
The latter choice was argued to be the correct prescription on the basis that a symmetry
of Minkowski space should not introduce physical defects into the geometry.  
A further justification comes from the explicit expressions (\ref{eqn:nullgeo}), (\ref{eqn:finitesuperrot})
for the finite superrotation, which reveal that the transformation is either multivalued or not surjective,
depending on the value of $\alpha$.  Instead, there is a principal domain that is either a
subset or a branched cover of Minkowski space that maps bijectively to the image Minkowski
space, and this principal domain determines the appropriate coordinate
ranges to be used in the superrotated geometries.  For the $z^\alpha$ transformation, 
this prescription in terms of the principal domain leads to the aforementioned periodicity of 
$\frac{2\pi}{\alpha}$ for $\phi$.  

The explicit superrotation formula was further exploited in section \ref{sec:embeddings} 
to create plots of the fixed-$(u,r)$ surfaces in Bondi coordinates, as well as the 
fixed-$(u,\rho)$ surfaces associated with Newman-Unti coordinates.  The surfaces were 
shown to intersect $\scri^+$ at singular points 
of the transformation, further emphasizing that 
the large $r$ expansion differs from the usual notion of an expansion near infinity
in the vicinity of these points.   The $r=0$ surface marking the boundary
of the Bondi coordinate patch was also plotted in
figure \ref{fig:r=0surf}, which clearly shows 
a nontrivial interior not covered by these coordinates.  Newman-Unti coordinates provide
a means for extending the geometry into this region, as described in section \ref{sec:extension}.   

This work has clarified several geometric aspects of finite superrotations, many of which
appear to have gone unnoticed previously.  Below, we discuss some 
physical implications, and comment on avenues for future work.

\subsection{Relaxing fixed-metric boundary conditions}
Changing the periodicity of $\phi$ eliminated 
the bulk cosmic string at the expense of introducing a defect in the induced metric on 
the celestial sphere.  
Standard treatments of null infinity impose a round sphere metric as $r\rightarrow\infty$ 
\cite{Bondi1962, Madler2016},\footnote{This is true in a fixed Bondi conformal gauge.
An invariant statement that does not refer to a conformal gauge is that a certain universal
structure consisting of a degenerate metric and null generator
modulo conformal rescalings is fixed on $\scri^+$  
as the boundary condition for the metric \cite{Ashtekar1981,Ashtekar2015}.} 
and this would preclude the defect sphere metrics.  
The requisite boundary conditions to accommodate the notion of superrotation action 
described in this
paper
must allow
for arbitrary punctures with arbitrary deficit angles, arising after acting with superrotations generated 
by functions $G(z)$ with poles and branches.  Note that adding punctures to the sphere
can be viewed as a changing the topology of $\scri^+$ away from the usual 
$S^2\times \mathbb{R}$ usually imposed for asymptotically flat spaces \cite{Wald1984}, 
so this represents a  generalization of asymptotic flatness that allows for 
cosmic string punctures at infinity.  One could further imagine approximating any 
metric on the celestial sphere by successively adding such defects. The conclusion
is that the boundary conditions should not restrict the allowed class of induced metrics at all
in order to allow for all possible superrotations.  

With relaxed boundary conditions on the celestial sphere metric, however, the symmetries 
of the spacetimes must be reconsidered.  The need for conformal isometries of the 
celestial sphere when determining the symmetry group of asymptotically flat spaces was tied
to the fixed round metric on the celestial sphere.  When this boundary condition is dropped, one
finds instead the $\text{Diff}(S^2)$ symmetry algebra appearing in
\cite{Campiglia2014a, Campiglia2015a}.   
Since such boundary conditions seem to be an unavoidable consequence of allowing
superrotations as symmetries, it appears much more natural to take $\text{Diff}(S^2)$ to be 
the fundamental symmetry of asymptotically flat spaces. 

To establish this proposal more rigorously, one would like to construct a phase space
that 
admits the $\text{Diff}(S^2)$ transformations as symmetries.  This requires finding a 
symplectic form that is appropriately conserved from $\scri^-$ to $\scri^+$ in the presence
of the weakened boundary conditions.  Additionally, finite, integrable charges must be 
constructed that generate the action of $\text{Diff}(S^2)$ on this phase space.  
See \cite{Compere2018, Flanagan2019} for recent work on this front.  We 
leave investigation into such a symplectic form to future work.

\subsection{Absence of defects}

Our focus in this paper was the presence or absence of string-like conical defects in the 
specific example of the $G(z) = z^\alpha$ superrotation.  This transformation maps the 
celestial sphere to a branched cover or subset of the sphere, and this extends to multivaluedness
or nonsurjectivity of the bulk transformation.  For more general superrotations, we should expect
multiple branch points which have similar behavior to that found for the $z^\alpha$ Bondi surfaces.  
An interesting example that demonstrates this point is given by the finite transformation
obtained by integrating a single Virasoro generator $Y^z = z^{-m+1}$.  This produces the 
finite transformation
\beq
G_\lambda(z) = e^{i \arg z}\Big(m\lambda e^{-im\arg z}+|z|^m\Big)^{1/m} 
\overset{*}{=} \Big(m \lambda + z^m\Big)^{1/m},
\eeq
where the second equality only holds when $\frac{-\pi}{m}<\arg z<\frac{\pi}{m}$, and is included
to show that this is indeed a holomorphic function over most of the complex plane.
Focusing on the case of $m>1$, this function has branch discontinuities extending radially from
the origin 
along 
$\arg z = \frac{2n+1}{m}\pi$, $n\in\mathbb{Z}$, out to branch points at $|z| = (m\lambda)^{1/m}$. 
The Schwarzian diverges at each of the branch points as well as at the origin where the 
branch cuts meet, and near these points we would again expect the spatial Bondi surfaces 
to intersect $\scri^+$.  

Note that the superrotation transformation given by (\ref{eqn:nullgeo}), (\ref{eqn:finitesuperrot})
manifestly is presented as a mapping from Minkowski space to itself, so that in particular the 
image spacetime is free of any defect.  The subtlety comes from the fact that this transformation
is not a diffeomorphism, since it can be multivalued and/or fail to be surjective.  
Defects only appear if one insists
that the superrotation map be bijective with its image.  

Finally, we mention the superrotation charges constructed in 
\cite{Barnich2011a, Flanagan:2015pxa, Compere:2016jwb}.  
For the Comp\`ere-Long metric (\ref{eqn:gCL}), these 
charges can be written in terms of integrals of $T_{AB}$ over a spatial Bondi surface at 
large $r$.  
It was claimed that nonzero values of these charges measure physical defects in the superrotated spacetimes.  
However, since we realized the superrotation as a transformation into regular Minkowski space and obtained the same metric, they cannot be detecting defects of the bulk spacetime.
Instead, the superrotation charges must be measuring some  geometric property of the embedding into Minkowski spacetime in the large $r$ limit.
A similar situation arises in asymptotically AdS${}_3$ gravity, where one can act with the Virasoro algebra on the CFT vacuum to obtain states containing ``boundary gravitons'' \cite{Brown:1986nw,Banados:1998gg}.
The resulting spacetime is empty AdS${}_3$, but with a different identification of the asymptotic boundary; the resulting stress tensor is nonzero, but this stress tensor simply measures the extrinsic curvature of the embedding.
We expect that the superrotation charges could similarly be expressed in terms of the geometry of the embedding into flat Minkowski spacetime.

\subsection{Newman-Unti gauge and hyperbolic slicing}

An interesting upshot of our investigations was a number of simplifications that 
occurred when using Newman-Unti coordinates instead of Bondi coordinates.  
The form of the supperrotated metric in Newman-Unti gauge is much simpler than in Bondi gauge,
since only the spatial metric components $g_{zz}$, $g_{z\bar z}$, and $g_{\bar z\bar z}$ are 
transformed according to equations (\ref{eqn:gABNU}), while the other metric components remain
the same as in Minkowski space in standard retarded coordinates.  
The Newman-Unti coordinates  are arguably more deeply tied to the geometry of Minkowski
space than are Bondi coordinates, 
since the radial coordinate is an affine parameter for the radial null geodesics.  Furthermore,
it was found that defining superrotations to preserve Newman-Unti gauge was equivalent to 
requiring the superrotation preserve hyperboloids at fixed Lorentzian distance to the origin.  
Some approaches to flat-space holography employ this hyperbolic slicing of Minkowski space
to apply aspects of the AdS/CFT dictionary on each slice 
\cite{DeBoer2003, Cheung2017}.  More recently, \cite{Ball2019}
noted that superrotations act in a natural way with respect to 
the hyperbolic slicing, and this is consistent with our findings on the simplifications afforded
by Newman-Unti gauge.  

 One advantage of the Newman-Unti coordinates is 
that they provide a means of constructing the hyperbolic slicing in a way that is tied to the asymptotic
boundary of the spacetime, rather than to a choice of origin in the interior.  In spacetimes with
matter or gravitational excitations, the Newman-Unti coordinates may give a more natural, 
boundary-anchored 
notion of the hyperbolic slicing, by defining the hyperboloids to be surfaces of fixed 
$\ell^2 = -u^2-2u \rho$.  In Minkowski space and its images under finite superrotations, this 
notion of hyperbolic slicing agrees with the usual one constructed relative to the origin.  
The Newman-Unti coordinates can be viewed as a flat space 
analog of the Fefferman-Graham expansion
employed in AdS, since in both cases the radial coordinate vector is a geodesic and the radial
coordinate is an affine parameter.  Newman-Unti gauge may therefore be helpful in 
future investigations into  flat-space holography.

\subsection{Action on twistor space}
The superrotation transformation used in this paper was constructed to preserve Bondi
gauge, which means that it maps one family of radial null geodesics to another.  Since 
any null geodesic can be transformed to a radial one in a retarded coordinate system using a 
translation, it follows that the superrotations in fact send any null geodesic to another null
geodesic.  This has the intriguing consequence that superrotations naturally act on the space
of projective null twistors, which is equivalent to the space of null geodesics of 4D Minkowski 
space
\cite{Penrose1986}.   
Twistors have appeared at various points in recent investigations into asymptotic symmetries
of Minkowski space.  The cosmic string construction used by Strominger and Zhiboedov 
\cite{Strominger:2016wns}
originally arose as a twistorial construction due to Nutku and 
Penrose \cite{Nutku1992},
and aspects of the relation between asymptotic symmetries and soft theorems have been
investigated from the viewpoint of ambitwistor strings \cite{Geyer2015}.  
An interesting proposal for future work would be to investigate the action of superrotations 
on twistor space, and see if it clarifies the role they play as asymptotic symmetries.  

\acknowledgments 
We thank Geoffrey Comp\`ere, Ted Jacobson, and Sabrina Pasterski for helpful discussions. 
We also are grateful to Sylvain Carrozza, Bianca Dittrich, Ma\"it\'e Dupuis,
and Florian Girelli,  
for organizing the 2019 PI-Waterloo Quantum Gravity Winter Retreat, 
where this work was initiated.  
Research at Perimeter Institute is supported in part by the Government of Canada through the Department of Innovation, Science and Economic Development Canada and by the Province of Ontario through the Ministry of Economic Development, Job Creation and Trade.

\appendix
\section{Solving for the finite superrotation}\label{sec::diffeo}
This appendix derives the explicit transformation 
of Minkowski space for a finite superrotation.  Such a formula was first derived
by Comp\`ere and Long in \cite{Compere2016b}.  An alternative derivation is provided 
below, that gives a more systematic derivation and presents the transformation in a way that 
is easy to determine where individual points map to.   

The bulk vector field generating the superrotation is defined by requiring that the Bondi gauge 
conditions are preserved along the flow.  These conditions are that the radial vector field 
$\partial_r^a$ be tangent to a null geodesic congruence, and that the determinant of the 
induced metric on surfaces of constant $(u,r)$ be equal to $-r^4 \gamma(z, \bar z)^2$.  
The general form of a vector field
whose flow preserves these conditions is given in 
\cite{Barnich2010c, Compere2018}, 
and those corresponding to a pure superrotation take the form
\begin{align}
\xi^A &= Y^A -\frac{u}{2 r}D^A (D\cdot Y) + \mathcal{O}(r^{-2}) \label{eqn:xiA} \\
\xi^u &= \frac{u}{2} D\cdot Y \label{eqn:xiu} \\
\xi^r &= -\frac{r}{2} D\cdot Y +\frac{u}{4} D_A D^A (D\cdot Y) + \mathcal{O}(r^{-1}). \label{eqn:xir}
\end{align}
$Y^A$ defines a local conformal Killing vector on the unit sphere, given explicitly in holomorphic
coordinates by $Y^A\partial_A = Y(z)\partial_z + \bar Y (\bar z) \partial_{\bar z}$.  Such a vector 
field satisfies the relation\footnote{This follows 
from a general identity for a conformal Killing vector $\zeta^a$
\beq
\nabla_b\nabla_c \zeta_d = R\indices{^a_b_c_d}\zeta_a +\frac{1}{d}\Big(g_{cd} \nabla_b
(\nabla \cdot \zeta) + g_{bd} \nabla_c (\nabla \cdot \zeta) - g_{bc} \nabla_d(\nabla \cdot \zeta) \Big)
\eeq
which, after taking a derivative and tracing leads to the identity
\beq \label{eqn:del2delz}
\nabla^2(\nabla \cdot \zeta) = -\frac2d\Big( \zeta^d\nabla_d R + R \nabla\cdot \zeta\Big).
\eeq
The unit 2-sphere has a Ricci scalar $R=2$, reducing (\ref{eqn:del2delz})
to the relation mentioned above.  
} 
$D_A D^A (D\cdot Y) = -2 (D\cdot Y)$, so that for a superrotation, $\xi^r$ becomes
\beq \label{eqn:xirsuperrot}
\xi^r = -\frac{r+u}{2} D\cdot Y + \mathcal{O}(r^{-1}).
\eeq

The subleading terms in equations (\ref{eqn:xiA}) and (\ref{eqn:xir})
depend on components of the metric in the Bondi coordinate system.  This causes the generators
to change along the flow of the vector field, complicating the problem of integrating this flow.
Fortunately, the terms explicitly
displayed in equations (\ref{eqn:xiA}-\ref{eqn:xir})  are all metric-independent,
allowing a solution for the flow to be found in the asymptotic region.  The asymptotic solution can 
then be extended to a full solution for the metric-dependent flow by imposing Bondi
gauge with a boundary condition set by the asymptotic solution.

\subsection{Asymptotic solution}
The solution to the flow equation can be expressed in the asymptotic region as a 
series in inverse powers of the initial radius $r_i$.  The asymptotic solution 
then takes the form
\begin{align}
u(\lambda) &= u_0(\lambda) +\mathcal{O}(r_i^{-1}) \\
r(\lambda) &= r_i r_{-1}(\lambda) + \mathcal{O}(r_i^{-1}) \\
z(\lambda) &= z_0(\lambda) + \frac{1}{r_i} z_1(\lambda) + \mathcal{O}(r_i^{-2}) \\
\bar z(\lambda) &= \bar z_0(\lambda) + \frac{1}{r_i} \bar z_1 (\lambda) + \mathcal{O}(r_i^{-2})
\end{align}
where $\lambda$ is the parameter along the flow.  
This expansion of the solution along with equations (\ref{eqn:xiA}-\ref{eqn:xir})
for the vector field produce the asymptotic differential equations to be solved,
\begin{align}
\dot u_0 &= \frac{u_0}{2} (D\cdot Y) [z_0, \bar z_0] \label{eqn:u0dot} \\
\dot z_0 &= Y(z_0) \label{eqn:z0dot} \\
\dot z_1 &= z_1 \partial Y[z_0] -\frac{ u_0}{2 r_{-1}} \frac{\bar \partial (D\cdot Y) [z_0,\bar z_0]}
{\gamma(z_0,\bar z_0)}, \label{eqn:z1dot}
\end{align}
and the equations involving $\bar z_0, \bar z_1$ are simply the complex conjugation of 
(\ref{eqn:z0dot}) and (\ref{eqn:z1dot}).  Note that the brackets in the expression
$(D\cdot Y)[z_0, \bar z_0]$ mean that first $D\cdot Y$ should be constructed as a function, and then
it should be evaluated on $(z_0, \bar z_0)$, and similarly
for $\bar \partial(D\cdot Y)[z_0,\bar z_0]$.  
The equation for $r_{-1}$  does not need to be solved separately due to an 
approximate conservation law that arises from the form of the generators given in 
(\ref{eqn:xiu}) and 
(\ref{eqn:xirsuperrot}).  The proper distance to the origin of Minkowski space, $-u^2 - 2 u r$, is 
preserved along the flow, up to $\mathcal{O}(r_i^{-1})$ terms.  This asymptotic conservation law
$u^2 + 2 u r = u_i^2 + 2 u_i r_i + \mathcal{O}(r_i^{-1})$ is then enough to determine $r_{-1}$  
in terms of other quantities,
\beq
r_{-1} = \frac{u_i}{u_0}.
\eeq

The solution to (\ref{eqn:z0dot}) is a finite holomorphic transformation $z_0 = G_\lambda(z_i)
\equiv G(z_i)$ obtained by integrating the flow of $Y(z)\partial_z$ in the complex plane.
This solution can then be used to integrate equation (\ref{eqn:u0dot}) once $D\cdot Y[z_0,\bar z_0]$
is determined
as a function of $\lambda$.  For this, it is useful to recall that $D\cdot Y$ appears 
as the infinitesimal change in the conformal factor of the unit sphere metric 
$q = 2\gamma(z,\bar z) dz d\bar z$, as captured by the 
conformal Killing equation
\beq \label{eqn:CKE}
\lie_Y q = (D\cdot Y) q.  
\eeq
The finite holomorphic transformation for the flow $z_i\rightarrow G_\lambda(z_i)$ is a conformal
isometry of the unit sphere metric, meaning the its pullback under the transformation is related to the 
original metric by a conformal factor, $G_\lambda^* q = \Omega_\lambda^2 q$.  This 
determines $\Omega_\lambda^2$ to be given by
\beq
\Omega_\lambda^2(z_i,\bar z_i) = \partial G \bar \partial \bar G 
\frac{\gamma(G,\bar G)}{\gamma(z_i,\bar z_i)}.
\eeq
However, equation (\ref{eqn:CKE}) implies that $\Omega_\lambda^2$ satisfies the 
differential equation
\beq
\frac{d}{d\lambda}\Omega_\lambda^2 = \Omega_\lambda^2 D\cdot Y[G_\lambda, \bar G_\lambda]
\eeq
with solution
\beq
\Omega_\lambda^2 = \exp\left[ \int_0^\lambda d \sigma\, (D\cdot Y)[G_\sigma, \bar G_\sigma]\right].
\eeq
This then leads to an explicit solution to (\ref{eqn:u0dot}),
\beq
u_0 = u_i \exp\left[\frac12\int_0^\lambda d\sigma \, (D\cdot Y)[G_\sigma,\bar G_\sigma] \right] 
 = u_i \Omega_\lambda
\eeq

The final equation determining the asymptotic solution is (\ref{eqn:z1dot}).  This is a first order,
linear differential equation of the form
\beq
\dot z_1 = a(\lambda) z_1 + b(\lambda)
\eeq
with 
\beq
a(\lambda) = \partial Y[G_\lambda], \qquad b(\lambda) = -\frac{u_i \Omega_\lambda^2}{2}
\frac{\bar\partial(D\cdot Y) [G_\lambda,\bar G_\lambda]}{\gamma(G_\lambda,\bar G_\lambda)}.
\eeq
The explicit solution for such an equation with initial condition $z_1(0)=0$ is \cite{Birkhoff}
\beq
z_1(\lambda)= e^{A(\lambda)} \int_0^\lambda d\sigma\, e^{-A(\sigma)} b(\sigma),\qquad
A(\lambda) = \int_0^\lambda d\sigma\, a(\sigma).
\eeq
The solution this produces for $z_1$ simplifies after some algebra to 
\beq
z_1 = - u_i \frac{\partial G_\lambda}{\gamma(z_i,\bar z_i)} \bar \partial \log \Omega_\lambda.
\eeq

\subsection{Extending to finite radius}
Having found the asymptotic solution to the superrotation flow, the extension
to finite radius is done by explicitly imposing the Bondi gauge conditions.  The first
is that lines of constant $(u_i, z_i, \bar z_i)$ map to a null geodesic.  An arbitrary null
geodesic in Minkowski space can be parameterized using Cartesian coordinates $(x^0, \vec{x})$ 
as
\begin{align}
x^0&=t \label{eqn:x0} \\
\vec{x} &= \vec b + c\, \hat \omega + t \hat \omega, \label{eqn:vecx}
\end{align}
where $t$ serves as a parameter along the geodesic, $\hat\omega$ is a spatial unit vector, and 
$\vec b \cdot \hat\omega = 0$.  The quantities $\vec b$, $c$, and $\hat\omega$ fully specify the null
ray, so they are determined entirely by the asymptotic solution and are independent of $r_i$.  
The function $t$ is the only quantity that depends on $r_i$, and this dependence will
be fixed by imposing the final Bondi gauge condition on the determinant of the spatial metric.  

Matching the large $r_i$ (and hence large $t$) limit of (\ref{eqn:x0}) and (\ref{eqn:vecx}) to 
the asymptotic solution found above fixes $\hat\omega$ to be the point on the unit sphere
corresponding to $G(z_i)$, explicitly
\beq\label{eqn:omega}
\hat\omega = \frac{1}{1+ G \bar G} 
\begin{pmatrix} G+ \bar G\\ -i(G-\bar G) \\ -1+ G\bar G\end{pmatrix}.
\eeq
Because $\vec b\cdot \hat \omega = 0$, $\vec b$ can be viewed as a tangent vector to the unit 
sphere at the point $(G(z_i), \bar G(\bar z_i))$.  Similarly, $z_1$ defines a small correction to 
the point $G(z_i)$ on the unit sphere, characterized in terms of the two-component vector 
$\begin{pmatrix} z_1\\\bar z_1 \end{pmatrix}$.  This vector determines $\vec b$, up to 
a rescaling by $\Omega$, through its pushfoward by the map $\hat \omega$ that embeds the 
unit sphere in $\mathbb{R}^3$,
\beq \label{eqn:bo*}
\vec b = \frac1\Omega \omega_* \begin{pmatrix} z_1\\\bar z_1\end{pmatrix}, 
\qquad b^i = \frac1\Omega\left( \frac{\partial\omega^i}{\partial G} z_1 
+ \frac{\partial\omega^i}{\partial \bar G} \bar z_1 \right).
\eeq
The final matching condition determines $c$ in terms of the asymptotic solution $u_0$,
\beq \label{eqn:c}
c = -u_i\, \Omega.
\eeq

All that remains is to determine how $t$ depends on the initial coordinates.  This is done
by computing the pulled back spatial metric for arbitrary $t$, and then determining
the functional dependence by imposing the final Bondi gauge condition on the spatial
metric determinant.  
In general, this procedure would produce a differential equation for $t$, but luckily all derivatives
of $t$ drop out once the spatial metric pullback is computed using 
the quantities $\vec b$, $\hat \omega$, and $c$ determined above.  

To show this, it is first useful to note that any vector in $\mathbb{R}^3$ that is orthogonal to 
$\hat\omega$ can be viewed as a tangent vector to the unit sphere, and hence can be realized as 
the pushfoward of a 2-component vector by $\omega_*$, as $\vec b$ was in (\ref{eqn:bo*}).  
One such vector is $\partial\hat\omega\equiv \frac{\partial}{\partial z_i} \hat \omega$ since 
$\hat \omega$ is identically of unit norm.  This leads to the relation $\partial\hat\omega = \omega_*
\begin{pmatrix} \partial G\\0\end{pmatrix}$, and similarly one finds $\bar\partial\hat\omega
= \omega_*\begin{pmatrix} 0 \\ \bar\partial\bar G \end{pmatrix}$.  When taking the dot product
between two vectors orthogonal to $\hat\omega$, one can instead represent it in terms 
of the corresponding 2-component vectors, with an inner product taken with respect to the 
sphere metric, $2\gamma(G,\bar G) dG d\bar G$.  A useful application of this procedure is in
computing $\hat\omega \cdot \partial\vec b = -\partial\hat\omega\cdot \vec b$, which gives
\beq\label{eqn:o.db}
\hat\omega\cdot \partial \vec b = -\frac1\Omega \gamma(G, \bar G) \partial G \bar z_1 
= u_i \partial\Omega.
\eeq
Then, forming the vector $\vec a\equiv \vec b + c \,\hat \omega$, equations (\ref{eqn:c}) and 
(\ref{eqn:o.db}) imply that $\hat\omega\cdot \partial \vec a = 0$, so that $\partial \vec a$ 
can be expressed in terms of a 2-component vector.  The 2-component vector can be 
computed straightforwardly but somewhat tediously, and this results in
\beq \label{eqn:da}
\partial\vec a = -\frac{u_i}{\Omega} \omega_*\begin{pmatrix} \partial G\, L \\
\frac{\bar\partial\bar G}{2 \gamma(z_i,\bar z_i)} \{ G(z_i), z_i\} \end{pmatrix} \equiv \omega_*\mu
\eeq
where 
\beq
L = \frac{1}{\gamma(z_i,\bar z_i)} \partial\log \Omega\, \bar\partial\log\Omega +\frac12(\Omega^2+1),
\eeq
and
\beq
\{ G(z_i) ,z_i\} = \partial\left(\frac{\partial^2 G}{\partial G}\right) 
- \frac12\left(\frac{\partial^2 G}{\partial G}\right)^2
\eeq
is the Schwarzian derivative of $G(z_i)$.  A similar argument for $\bar\partial \vec a$
yields
\beq \label{eqn:dbara} 
\bar\partial\vec a = -\frac{u_i}{\Omega} \omega_* 
\begin{pmatrix} \frac{\partial G}{2 \gamma(z_i,\bar z_i)} \{\bar G(\bar z_i), \bar z_i\} \\
\bar \partial\bar G\, L \end{pmatrix}  = \omega_*\nu
\eeq

Given these relations, the next step is to find an expression for the pulled back spatial metric.  
The image of the diffeomorphism is expressed in (\ref{eqn:x0}) and (\ref{eqn:vecx}) in 
Cartesian coordinates on Minkowski space, so the metric being pulled back is simply
$\eta_{\mu\nu} = \diag(-1,1,1,1)$.  Letting $z^A = (z_i, \bar z_i)$ denote the coordinates
on the constant $(u_i, r_i)$ surfaces and using that $\hat\omega$ is a unit vector
and that $\partial \vec a \cdot \hat\omega = \bar\partial\vec a \cdot \hat \omega = 0$, one obtains
the spatial metric
\beq\label{eqn:gAB3}
g_{AB} = \left(\partial_A \vec a + t \partial_A\hat\omega \right) 
\cdot \left(\partial_B \vec a + t \partial_B \hat \omega\right),
\eeq
where, as advertised, all derivatives of $t$ have dropped out.  
Since both $\partial_A \vec a$ 
and $\partial_A \hat \omega$ are orthogonal to $\hat\omega$, it is useful to express
this equation in terms of 2-component objects.  The pair $\mu$ and $\nu$  
from (\ref{eqn:da}) and (\ref{eqn:dbara}) can be combined into a $2 \times 2$ matrix,
\beq
A = \begin{pmatrix} \mu & \nu \end{pmatrix} 
\eeq
which pushes forward to $\partial_A \vec a = \begin{pmatrix}
\partial \vec a & \bar\partial\vec a\end{pmatrix}$, and $\partial_A\hat\omega$ is similarly
represented as the pushfoward of 
\beq
W = \begin{pmatrix} \partial G & 0 \\ 0 & \bar\partial\bar G \end{pmatrix}.
\eeq
The dot product in terms of the 2-component vectors is taken using the sphere metric $q_G$, 
represented as a matrix by 
\beq
q_G = \begin{pmatrix} 0 & \gamma(G,\bar G) \\ \gamma(G,\bar G) & 0 \end{pmatrix}.
\eeq
In terms of these matrices, equation (\ref{eqn:gAB3}) for $g_{AB}$ can instead be expressed
as the matrix product
\beq
g_{AB} = \left(A + t W\right)^T \cdot q_G \cdot \left(A + t W \right).
\eeq

This expression greatly simplifies the computation of the determinant of $g_{AB}$,
and the Bondi gauge condition becomes
\begin{align}
\det g_{AB} &= \left[\det(A + t W)\right]^2 \det q_G = -r_i^4 \gamma(z_i,\bar z_i)^2 \\
\det(A + t W) &= r_i^2 \frac{\gamma(z_i,\bar z_i)}{\gamma(G,\bar G)} 
= \frac{r_i^2\partial G\bar\partial\bar G}{\Omega^2}. 
\end{align} 
The result is a quadratic equation for $t$,
\beq
t^2 -\frac{2u_i L}{\Omega} t +\left(\frac{u_i L}{\Omega}\right)^2 - \frac{r_i^2 + u_i^2 V}{\Omega^2} = 0
\eeq
where
\beq
V = \frac{1}{4 \gamma(z_i,\bar z_i)^2} \{G(z_i),z_i\}\{\bar G(\bar z_i), \bar z_i\},
\eeq
and the solution with the correct large $r_i$ behavior is 
\beq \label{eqn:t}
t = \frac{1}{\Omega} \left( u_i L + \sqrt{r_i^2 + u_i^2 V}\right).
\eeq
This completes the derivation of the action of the superrotation on Minkowski space, and the 
final transformation is presented in equations (\ref{eqn:nullgeo}) and (\ref{eqn:finitesuperrot}).

It is relatively straightfoward to check that the coordinate transformation 
(\ref{eqn:x0}), (\ref{eqn:vecx}) with $t$, $\vec b$, $c$ and $\hat\omega$ given by 
(\ref{eqn:t}), (\ref{eqn:bo*}), (\ref{eqn:c}) and (\ref{eqn:omega}) reproduces the 
form of the metric derived by Comp\`ere and Long when specialized to 
a pure superrotation transformation.  This confirms that the coordinate transformation
agrees with the one presented in \cite{Compere2016b, Compere:2016jwb}.  

To obtain the transformation in Newman-Unti gauge, all steps are the same except for how we 
solve for $t$ in terms of the radial coordinate $\rho$.  It must be related to $\rho$ by an
affine tranformation $t = f(u_i,z_i,\bar z_i)+ g(u_i, z_i,\bar z_i)\rho$, and demanding that 
$g^{\alpha u} = \delta^\alpha_\rho$, as required by the gauge condition, fixes $g = \frac1\Omega$.  
For the final condition, we need to compute $\partial_\rho \log \det g_{AB}$,
\begin{align}
\partial_\rho \log\det g_{AB} &= 2 \partial_\rho \det(A + t W) = 
2 \partial_\rho\log\left(t^2 -\frac{2u_i L}{\Omega} t + \left(\frac{u_i L}{\Omega}\right)^2 
-\frac{u^2 V}{\Omega^2} \right) \nonumber \\
&=\frac2\rho \left(1 + \frac{u_iL-\Omega f}{\rho} + \mathcal{O}(\rho^{-2}) \right).
\end{align}
The Newman-Unti gauge choice is to set the first subleading term in this expansion to zero
\cite{Barnich2012}, 
and this determines $f = \frac{u_i L}{\Omega}$.  Hence the Newman-Unti superrotation in defined 
with 
\beq
t = \frac{1}{\Omega}(u_i L + \rho). 
\eeq
After computing the metric pullback with this transformation, we find 
\begin{subequations}
\label{eqn:gNUcomp}
\begin{align}
g_{uu} &= g_{ur} = -1 \\
g_{zz} &= -\rho u\{G,z\} \\
g_{\bar z\bar z} &= -\rho u \{\bar G,\bar z\}\\
g_{z\bar z}&= \gamma(z,\bar z)(\rho^2 + u^2 V),
\end{align}
\end{subequations}
with all other metric components vanishing.

\bibliographystyle{JHEPthesis}
\bibliography{football}

\end{document}